\documentclass[]{spie}  

\usepackage{amsmath,amsfonts,amssymb}
\usepackage{graphicx}
\usepackage[colorlinks=true, allcolors=blue]{hyperref}

\title{The Roman Coronagraph Community Participation Program: pre-launch reference star list and impact of reference star properties on post-processing performance\footnote{This paper includes data gathered with the 6.5 meter Magellan Telescopes located at Las Campanas Observatory, Chile.}}

\author[a]{Justin Hom}
\author[a]{Schuyler G. Wolff}
\author[b]{Jessica Gersh-Range}
\author[a]{Ramya M. Anche}
\author[c]{Vanessa P. Bailey}
\author[d]{Jean-Philippe Berger}
\author[e,f]{Beth A. Biller}
\author[g]{Wolfgang Brandner}
\author[h]{Marah Brinjikji}
\author[g]{Ga{\"e}l Chauvin}
\author[i]{David R. Ciardi}
\author[i]{Catherine A. Clark}
\author[a]{Laird M. Close}
\author[j]{Robert J. De Rosa}
\author[k]{Sarah Deveny}
\author[a]{Warren B. Foster}
\author[l]{Julien H. Girard}
\author[i]{Alexandra Z. Greenbaum}
\author[a,m,n,o]{Olivier Guyon}
\author[a,p]{Sebastiaan Y. Haffert}
\author[q]{Alexander D. Hedglen}
\author[r]{Steve B. Howell}
\author[a]{Parker T. Johnson}
\author[a,m]{Maggie Y. Kautz}
\author[m]{Jay K. Kueny}
\author[o,s]{Masayuki Kuzuhara}
\author[d]{Manon Lallement}
\author[t]{Rico Landman}
\author[k]{Colin Littlefield}
\author[d]{Jean-Baptiste Le Bouquin}
\author[a]{Jialin Li}
\author[m]{Joshua Liberman}
\author[u]{Joseph D. Long}
\author[a,n]{Miles Lucas}
\author[m]{Jennifer Lumbres}
\author[v,w]{Bruce Macintosh}
\author[a]{Jared R. Males}
\author[c]{Eric Mamajek}
\author[p]{Matthijs Mars}
\author[m]{Eden A. McEwen}
\author[x]{Avalon L. McLeod}
\author[y]{Maxwell A. Millar-Blanchaer}
\author[o,s]{Toshiyuki Mizuki}
\author[z]{Sophie Noiret}
\author[m]{Tiffany Nguyen}
\author[aa]{Lauren Schatz}
\author[a]{Nicholas T. Schragal}
\author[p]{Adam K. Taras}
\author[p]{Elena Tonucci}
\author[m]{Katie Twitchell}
\author[a]{Kyle Van Gorkom}
\author[g]{Macarena C. Vega-Pallauta}
\author[bb,cc]{Jason J. Wang}
\author[y]{Jingwen Zhang}
\affil[a]{Steward Observatory and Department of Astronomy, University of Arizona, 933 N Cherry Avenue, Tucson, AZ 85721, USA}
\affil[b]{DM Telescopes LLC, Raleigh, NC, USA}
\affil[c]{NASA Jet Propulsion Laboratory, California Institute of Technology, Pasadena, CA 91109, USA}
\affil[d]{Institute for Astronomy, University of Edinburgh, Royal Observatory, Edinburgh EH9 3HJ, UK}
\affil[e]{Centre for Exoplanet Science , University of Edinburgh, Edinburgh, EH9 3FD, UK}
\affil[f]{Universit{\'e} Grenoble Alpes, CNRS, IPAG, 38000 Grenoble, France}
\affil[g]{Max Planck Institute for Astronomy, K{\"o}nigstuhl 17, 69117 Heidelberg, Germany}
\affil[h]{Department of Physics and Astronomy, University of Notre Dame, Nieuwland Science Hall, Notre Dame, IN 46556, USA}
\affil[i]{NASA Exoplanet Science Institute, IPAC, California Institute of Technology, Pasadena, CA 91125, USA}
\affil[j]{European Southern Observatory, Alonso de C\'{o}rdova 3107, Vitacura, Casilla 19001, Santiago, Chile}
\affil[k]{Bay Area Environmental Research Institute, Moffett Field, CA 94035, USA}
\affil[l]{Space Telescope Science Institute, 3700 San Martin Drive, Baltimore, MD 21218, USA}
\affil[m]{Wyant College of Optical Sciences, The University of Arizona, Tucson, Arizona, USA}
\affil[n]{Subaru Telescope, National Observatory of Japan, National Institutes of Natural Sciences, 650 N. A'ohoku Place, Hilo, Hawai'i}
\affil[o]{Astrobiology Center of NINS, 2-21-1, Osawa, Mitaka, Tokyo, 181-8588, Japan}
\affil[p]{Leiden Observatory, Leiden University, PO Box 9513, 2300 RA Leiden, The Netherlands}
\affil[q]{Northrop Grumman Corporation, 600 South Hicks Road, Rolling Meadows, Illinois}
\affil[r]{NASA Ames Research Center, Moffett Field, CA 94035, USA}

\affil[s]{National Astronomical Observatory of Japan, 2-21-2, Osawa, Mitaka, Tokyo, 181-8588, Japan}
\affil[t]{NOVA, Netherlands Research School for Astronomy, P.O. Box 9513, NL-2300 RA Leiden, The Netherlands}
\affil[u]{Center for Computational Astrophysics, Flatiron Institute, 162 5th Avenue, New York, New York}
\affil[v]{Department of Astronomy and Astrophysics, University of California, Santa Cruz, Santa Cruz, CA 95064, USA}
\affil[w]{University of California Observatories, 1156 High Street, Santa Cruz, CA 95064, USA}
\affil[x]{Draper Laboratory, 555 Technology Square, Cambridge, Massachusetts}
\affil[y]{Department of Physics, University of California, Santa Barbara, Santa Barbara, CA, USA}
\affil[z]{Aix Marseille Univ, CNRS, CNES, LAM, Marseille, France}
\affil[aa]{Starfire Optical Range, Kirtland Air Force Base, Albuquerque, New Mexico}

\affil[bb]{Department of Physics and Astronomy, Northwestern University, 2145 Sheridan Road, Evanston, IL 60208-3112, USA}
\affil[cc]{Center for Interdisciplinary Exploration and Research in Astrophysics, 1800 Sherman Ave, Northwestern University, Evanston, IL 60201, USA}

\authorinfo{Further author information: (Send correspondence to J. Hom)\\J. Hom: E-mail: jrhom@arizona.edu}

\begin{document} 
\maketitle

\begin{abstract}
The upcoming Roman Coronagraph will be the first high-contrast instrument in space capable of high-order wavefront sensing and control technologies, a critical technology demonstration for the proposed Habitable Worlds Observatory (HWO) that aims to directly image and characterize habitable exoEarths. The nominal Roman Coronagraph observing plan involves alternating observations of a science target and a bright, nearby reference star for both wavefront calibration and reference differential imaging post-processing. Reference star criteria for the most demanding coronagraph mode are restrictive, limiting the sample to only 40 candidates for which thorough observational vetting is needed to assess their suitability. Reference star properties such as resolved diameters, presence of circumstellar dust, and close point sources may also have more subtle impacts on post-processing efficacy that may inhibit final contrast performance. In this work, we describe the current progress of the CoronaGraph Instrument Reference stars for Exoplanets (CorGI-REx) observing campaign, a $300+$-hour observing campaign that utilizes instruments from around the world to vet reference stars for high-order wavefront control suitability. We will present the pre-launch list of reference star candidates being utilized for the Roman Coronagraph Observation Phase constructed from a thorough analysis of high contrast and interferometric observations. We will also present the results of simulations investigating the impact of reference star resolved diameters and companions on post-processing performance. We conclude by discussing the importance of reference star selection for scheduling observations and optimizing contrast performance for the Roman Coronagraph along with implications for HWO coronagraph operations. 
\end{abstract}


\section{INTRODUCTION}
\label{sec:intro}  
The Roman Coronagraph is a critical technology pathfinder for the Habitable Worlds Observatory (HWO) with plans to demonstrate high-order wavefront sensing and control technologies in space for the first time\cite{mennesson2020}. High-order wavefront sensing and control (HOWFSC) and reference differential imaging (RDI) require observations of bright reference stars prior to observing science targets. These bright reference stars must follow strict criteria for optimizing scientific performance and ultimately drive the scheduling of a science program. Even after a list of suitable reference stars has been constructed, there may be other factors that could ultimately impact final science performance, particularly in the achievable post-processing gain from RDI. 

Hom et al.\cite{hom2026} introduced the criteria for Roman Coronagraph reference stars selected for optimal performance with the Hybrid-Lyot Coronagraph Narrow Field-of-View (HLC-NFOV) imaging mode. Hom et al.\cite{hom2026} presented the initial list of probable reference star candidates that satisfied the criteria, including having $V<3$, uniform stellar disk diameter $UDD_V<2$ mas, and having no companions (bound or background) that would inhibit HOWFSC or RDI performance. Reference stars must also be within 5$^{\circ}$ of observatory pitch angle from the science target at the time of observation to avoid severe contrast-degrading thermal variations. Despite a thorough literature and catalog search, the presence of companions could not be immediately ruled out for the initial list of reference star candidates, necessitating the assignment of confidence ``rankings" for use in a Roman Coronagraph observation. To confirm whether or not the initial list of reference star candidates are suitable for Roman Coronagraph operations, the CoronaGraph Instrument Reference stars for Exoplanets (CorGI-REx) observing campaign was initiated. The campaign seeks to assess archival/literature sources in addition to conducting new direct observations to reject the presence of stellar companions around the initial list of reference star candidates. Hom et al.\cite{hom2026} presented the results of moderate contrast ($\Delta mag \leq 8$) adaptive optics (AO) imaging and speckle interferometry of most of the reference star candidates. To be confirmed as suitable, however, the presence of stellar companions must be rejected to contrast limits of $\Delta mag \leq 17.5$. To reach these detection limits, the CorGI-REx campaign has been utilizing high-angular resolution optical interferometry and high-contrast imaging with several facilities around the world.

The initial list presented in Hom et al.\cite{hom2026} only considered stars optimized for the HLC-NFOV mode. However, the Roman Coronagraph also hosts a few ``best-effort" modes, including wide-field of view imaging and slit spectroscopy with shaped-pupil coronagraph masks (SPC-WFOV and SPC-SPEC respectively). With properties distinct from the HLC-NFOV mode, these best-effort modes also have unique criteria for ideal reference star selection.

In this work, we present the current observational results of the CorGI-REx campaign. In Section \ref{sec:newreflist}, we describe the criteria for reference star selection in the best effort SPC-WFOV and SPC-SPEC modes. In Section \ref{sec:observations}, we describe the current status of CorGI-REx observations of all reference star candidates, including the results from recent AO imaging and speckle interferometry campaigns. In Section \ref{sec:rankings}, we describe the process of re-evaluating reference star rankings based on recent observational results and new criteria that may impact suitability. In Section \ref{sec:discussion}, we describe the current impact of the reference star list on scheduling of science programs. We also present the results of various Roman Coronagraph simulations investigating the impact of certain reference star properties on RDI post-processing performance. In Section \ref{sec:summary}, we summarize the results of our findings and describe the next actions of the CorGI-REx campaign.

\section{New Reference Star List for Shaped-Pupil Coronagraph Modes} \label{sec:newreflist}
The original list of reference stars from Hom et al.\cite{hom2026} satisfy criteria optimized for Roman Coronagraph observations with the HLC-NFOV Band 1 mode. Aside from needing to be single star systems, the reference star candidates must be bright in $V$-band ($V<3$) for efficient HOWFSC times and have small resolved stellar diameters ($UDD_V < 2$mas) for mitigating low-order aberrations similar to pointing jitter\cite{krist2023}. This current reference star list, however, is not well-optimized for Roman Coronagraph observing modes using shaped-pupil coronagraph masks (SPC-SPEC Band 3 and SPC-WFOV Bands 1 and 4)\footnote{Band 1 $\lambda_C = 575$nm, Band 3 $\lambda_C = 730$nm, Band 4 $\lambda_C = 825$nm}. Given the lower system throughput and lower quantum efficiency of the science camera particularly at Bands 3 and 4, reference stars must be even brighter to avoid prohibitively long HOWFSC times (we choose a threshold of $I<1.25$). While there is an even smaller number of $I<1.25$ than $V<3$ stars, the SPC modes are not as sensitive to low order aberrations as the HLC mode. This allows stars with much larger resolved diameters to be used with minimal performance impact\cite{krist2023}. Similar to the HLC-optimized list, SPC-optimized reference stars must also not have companions that could degrade achievable post-processed contrast.

The construction of the SPC-optimized red reference star list began by identifying $I<1.25$ stars. Similar to the HLC-optimized list in Hom et al.\cite{hom2026}, we queried the SB9 catalog\cite{pourbaix2004}, \textit{Gaia}, and all literature sources for the candidates to remove all stars with confirmed companions. For the 26 stars that remain, there are varying degrees of evidence for and against these stars having stellar companions. With the first SPC Band 3 and 4 observations planned for Summer 2027, there is insufficient time to complete as deep of a vetting program as the primary HLC list. Ground-based observing time, however, has been secured using adaptive optics imaging with Palomar/PHARO-AO and speckle interferometry with Gemini-North/`Alopeke and Gemini-South/Zorro. These observations allow us to vet the red reference star list at more moderate contrasts and increase confidence in their use for Roman Coronagraph operations. Additionally, as a majority of the red reference star list are red giant branch stars, multiplicity is not as high of a concern\cite{badenes2018} compared to the HLC-optimized primary reference star list. The list of Band 3 and 4 reference star candidates and their properties are shown in Table \ref{tab:redlist_props}. It is also worth mentioning that these reference stars could also be used for SPC-WFOV imaging in Band 1 in combination with the existing HLC-NFOV reference stars and the reserve list described in Hom et al.\cite{hom2026}.


\begin{table}[ht]
\caption{List of new Band 3 and 4 reference star candidates. $\alpha$ Aql, $\beta$ Ori, and $\delta$ CMa also qualify as Band 3 and 4 reference stars but were also included as part of the ``reserve" reference star list in Hom et al.\cite{hom2026}. $UDD_I$ and $LDD$ are sourced from Bourges et al.\cite{bourges2014} except where indicated. Distances are calculated from \textit{Gaia} DR3 parallaxes \cite{gaiadr3}.} 
\label{tab:redlist_props}
\begin{center}       
\begin{tabular}{|l|l|l|l|l|l|l|l|l|}
\hline
Name & HD & HIP & SpTy & $V$ & $I$ & $UDD_I$ [mas] & $LDD$ [mas] & D [pc] \\ \hline
$\alpha$ Boo	&	124897	&	69673	&	K1.5IIIFe-0.5\cite{keenan1989}	&	-0.05	&	-1.68	&	20.00\cite{mozurkewich2003}	&	21.37	&	11.26	\\
$\beta$ Gru	&	214952	&	112122	&	M4.5III\cite{keenan1989}	&	2.11	&	-1.57	&	...	&	...	&	54.26	\\
$\gamma$ Cru	&	108903	&	61084	&	M3.5III\cite{keenan1989}	&	1.64	&	-1.44	&	26.73	&	28.05	&	27.15	\\
$\alpha$ Tau	&	29139	&	21421	&	K5+III\cite{keenan1989}	&	0.86	&	-1.31	&	21.73	&	22.88	&	20.43	\\
$\alpha$ Car	&	45348	&	30438	&	A9II\cite{gray1989F}	&	-0.74	&	-1.13	&	6.82	&	7.04	&	94.79	\\
$\beta$ Peg	&	217906	&	113881	&	M2.5II-III\cite{keenan1989}	&	2.42	&	-0.40	&	18.81	&	19.74	&	60.10	\\
$\beta$ And	&	6860	&	5447	&	M0+IIIa\cite{keenan1989}	&	2.05	&	-0.19	&	13.67	&	14.24	&	60.53	\\
$\beta$ Gem	&	62509	&	37826	&	K0IIIb\cite{keenan1989}	&	1.14	&	-0.11	&	7.06	&	7.39	&	10.36	\\
$\mu$ Gem	&	44478	&	30343	&	M3IIIab\cite{keenan1989}	&	2.87	&	-0.08	&	14.69	&	15.41	&	71.02	\\
$\rho$ Per	&	19058	&	14354	&	M4+IIIa\cite{keenan1989}	&	3.39	&	-0.03	&	15.38	&	16.14	&	94.34	\\
$\alpha$ Cet	&	18884	&	14135	&	M1.5IIIa\cite{keenan1989}	&	2.53	&	0.02	&	12.27\cite{mozurkewich2003}	&	13.24	&	76.39	\\
$\lambda$ Vel	&	78647	&	44816	&	K4Ib\cite{keenan1989}	&	2.21	&	0.03	&	11.32	&	11.91	&	166.94	\\
$\alpha$ Hya	&	81797	&	46390	&	K3IIIa\cite{keenan1989}	&	1.97	&	0.16	&	8.42	&	8.84	&	55.28	\\
$\beta$ UMi	&	131873	&	72607	&	K4-III\cite{keenan1989}	&	2.08	&	0.21	&	9.65	&	10.15	&	40.14	\\
$\gamma$ Dra	&	164058	&	87833	&	K5III\cite{keenan1989}	&	2.23	&	0.23	&	8.94	&	9.42	&	47.30	\\
$\delta$ Oph	&	146051	&	79593	&	M0.5III\cite{keenan1989}	&	2.75	&	0.43	&	9.81	&	10.24	&	49.00	\\
$\alpha$ Ari	&	12929	&	9884	&	K2-IIIbCa-1\cite{keenan1989}	&	2.01	&	0.54	&	7.26	&	7.62	&	20.18	\\
$\epsilon$ Peg	&	206778	&	107315	&	K2Ib-II\cite{keenan1989}	&	2.39	&	0.58	&	6.97\cite{mozurkewich2003}	&	7.46	&	211.42	\\
$\theta$ Cen	&	123139	&	68933	&	K0-IIIb\cite{keenan1989}	&	2.05	&	0.76	&	5.30	&	5.31	&	18.03	\\
$\epsilon$ Sco	&	151680	&	82396	&	K1III\cite{gray2006}	&	2.29	&	0.83	&	5.65	&	5.93	&	19.54	\\
$\alpha$ Cas	&	3712	&	3179	&	K0-IIIa\cite{keenan1989}	&	2.23	&	0.85	&	5.53	&	5.79	&	70.97	\\

\hline 
\end{tabular}
\end{center}
\end{table}

\section{Recently Obtained Observations} \label{sec:observations}
\subsection{Reference Star Vetting with Direct AO Imaging and Speckle Interferometry}
\begin{figure}
    \centering
    \includegraphics[width=\linewidth]{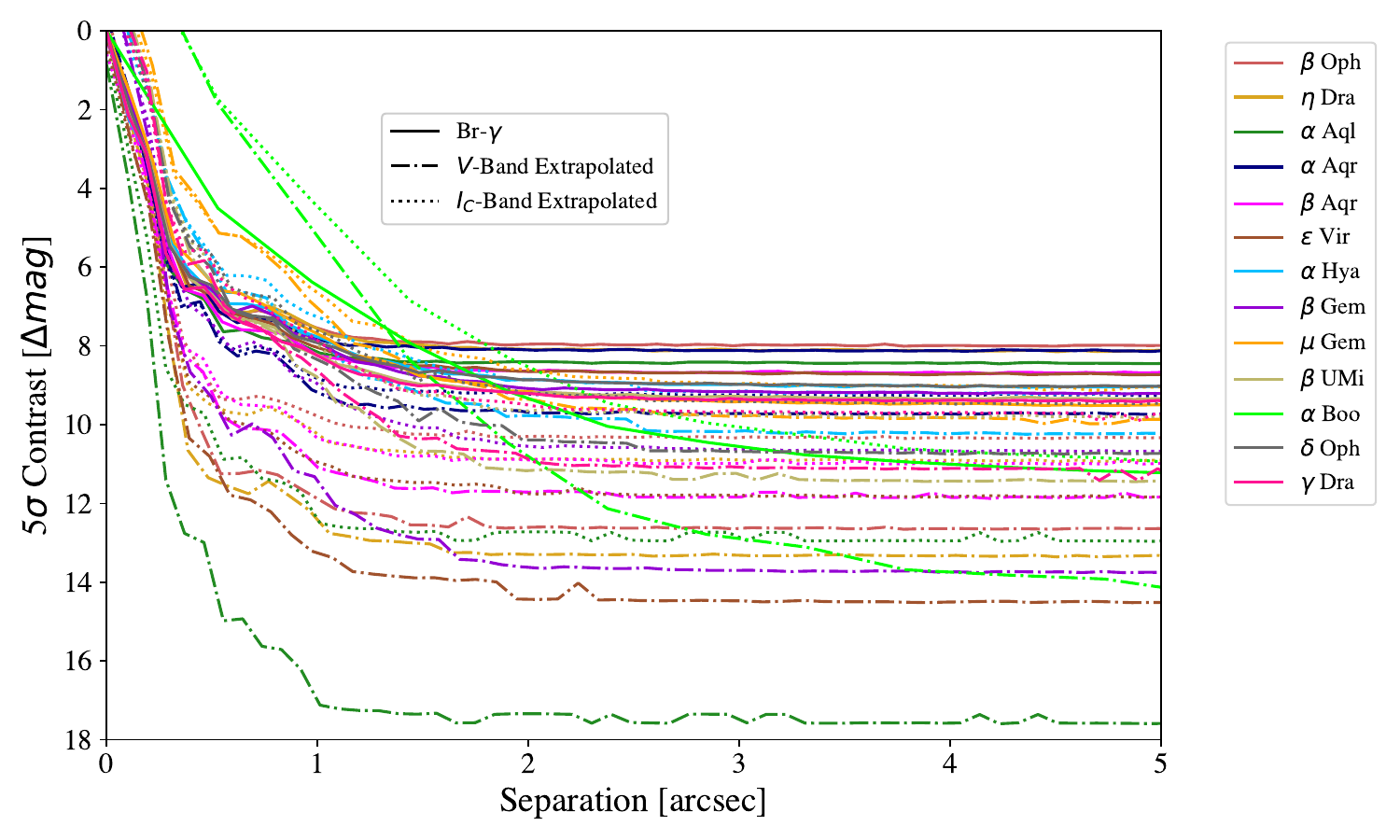}
    \caption{$5\sigma$ detection limit curves from PHARO observations of five reserve reference stars. The observations prioritized larger diameter stars that could be used for the CGI WFOV and SPEC modes. The dashed lines are extrapolations determined from the observed detection limits and the stellar distances assuming sensitivity to main sequence companions using colors from Pecaut \& Mamajek \cite{pecaut2013}.}
    \label{fig:PHARO_curves}
\end{figure}

We used the Palomar High Angular Resolution Observer (PHARO \cite{hayward2001}) at Palomar Observatory to observe some CorGI-REx candidate stars in the 2025B-2026A semesters (PI: C. Clark). We used the narrow Br$\gamma$ filter in combination with the narrow H2 filter in the grism wheel to avoid saturation. The data were collected in a cross dither pattern and reduced as described in Hom et al \cite{hom2026}. The summary of observations is reported in Table \ref{tab:PHAROobs}, and the $5\sigma$ detection limit curves are shown in Figure \ref{fig:PHARO_curves}.

\begin{figure}
    \centering
    \includegraphics[width=1\linewidth]{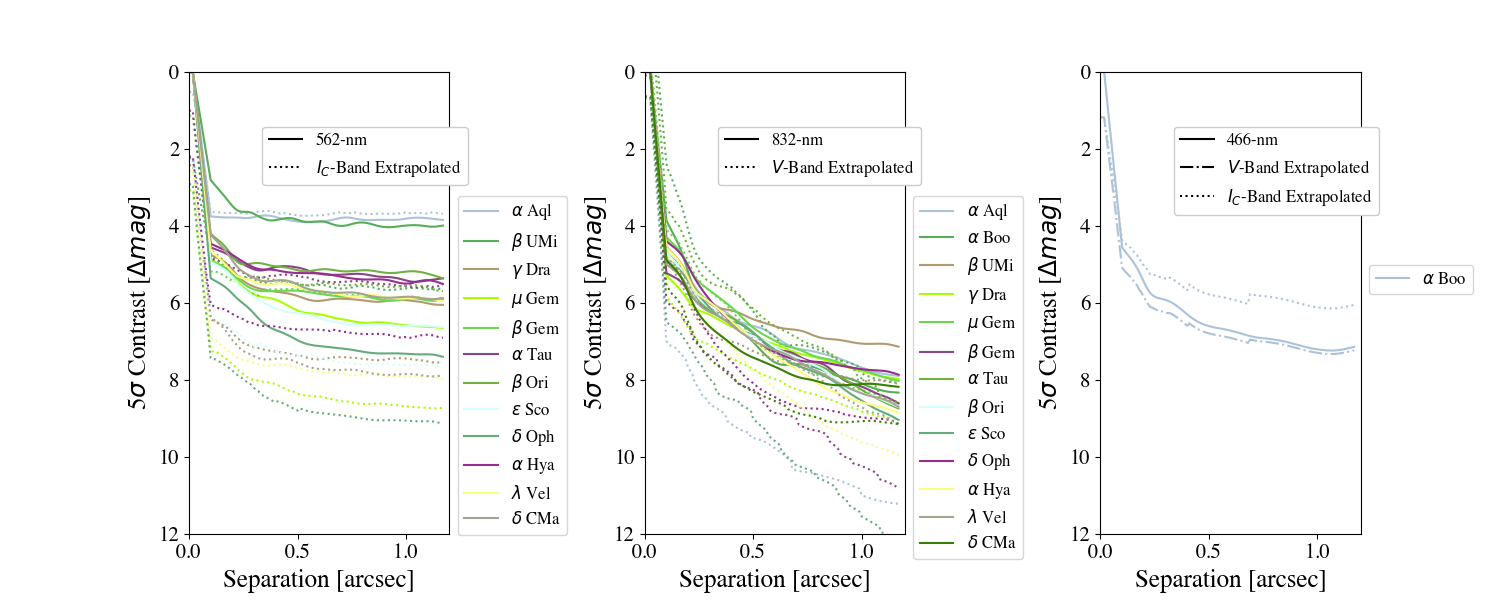}
    \caption{$5\sigma$ detection limit curves from speckle observations of Band 3 and 4 reference stars. The dashed lines are extrapolations determined from the observed detection limits and the stellar distances assuming sensitivity to main sequence companions using colors from Pecaut \& Mamajek \cite{pecaut2013}.}
    \label{fig:speckle_curves}
\end{figure}

We also performed speckle interferometry using the twin dual-channel imagers `Alopeke and Zorro \cite{scott2021} installed at the Gemini-North and Gemini-South observatories respectively. Observations were performed over the 2026A semester (PIDs: GN-2026A-FT-107, GS-2026A-FT-208; PI: J. Hom). The observations are summarized in Table \ref{tab:speckleobs}. We use the standard `Alopeke and Zorro optical wavelength filters (562-nm and 832-nm) for most observations. For $\alpha$ Boo we use the 466-nm blue channel filter instead of the 562-nm to avoid saturation. Targets were observed with several thousand short exposures over $\sim$10 minutes of observation time. Each observation is preceded or followed by an observation of a single PSF calibrator. Fourier analysis is used on the exposures to produce reconstructed images, robust $5\sigma$ contrast limits in each bandpass, and the flux and location of candidate companions \cite{horch2011,howell2011}. Observations are summarized in Table \ref{tab:speckleobs} and detection limits in each of the observed filters are shown in Figure \ref{fig:speckle_curves}. 
No new companions were detected from any AO imaging or speckle interferometry observations, encouraging their use in Roman Coronagraph observations.

\begin{table}[ht]
\caption{Summary of reference star observations taken with Palomar/PHARO. Contrast values are reported at 0."6.} 
\label{tab:PHAROobs}
\begin{center}       
\begin{tabular}{|l|l|l|l|}
\hline
Name	&	Date [UT]	&	$t_{int}$ [s]	&	$5\sigma$ Contrast [$\Delta mag$]	\\ \hline
$\beta$ Oph	&	20250804	&	12.6	&	7.18	\\
$\eta$ Dra	&	20250804	&	25.2	&	7.12	\\
$\alpha$ Aql	&	20250909	&	12.6	&	7.59	\\
$\alpha$ Aqr	&	20250909	&	25.2	&	7.41	\\
$\beta$ Aqr	&	20250909	&	51.3	&	7.59	\\
$\epsilon$ Vir	&	20260226	&	51.3	&	7.16	\\
$\beta$ UMi	&	20260226	&	12.6	&	7.31	\\
$\alpha$ Boo	&	20260226	&	12.6	&	4.50	\\
$\delta$ Oph	&	20260226	&	12.6	&	7.06	\\
$\gamma$ Dra	&	20260226	&	12.6	&	7.12	\\
$\alpha$ Hya	&	20260227	&	12.6	&	6.93	\\
$\beta$ Gem	&	20260227	&	12.6	&	7.15	\\
$\mu$ Gem	&	20260227	&	12.6	&	6.67	\\
\hline
\end{tabular}
\end{center}
\end{table}

\begin{table}[ht]
\caption{Summary of reference star observations taken with speckle interferometry instruments `Alopeke and Zorro. Contrast values are reported at 0."6.} 
\label{tab:speckleobs}
\begin{center}       
\begin{tabular}{|l|l|l|l|l|l|}
\hline
Name	&	Date [UT]	&	$t_{int}$ [s]	&	562-nm $5\sigma$ Contrast	&	832-nm $5\sigma$ Contrast	&	466-nm $5\sigma$ Contrast	\\
	&	[UT]	&	[s]	&	[$\Delta mag$]	& [$\Delta mag$]	&	[$\Delta mag$]	\\\hline
$\alpha$ Aql	&	20260402	&	180	&	3.88	&	6.86	&	...	\\
$\alpha$ Boo	&	20260402	&	186	&	...	&	7.54	&	6.72	\\
$\beta$ UMi	&	20260402	&	180	&	3.88	&	6.61	&	...	\\
$\gamma$ Dra	&	20260402	&	200	&	5.95	&	7.07	&	...	\\
$\mu$ Gem	&	20260402	&	180	&	6.29	&	6.98	&	...	\\
$\beta$ Gem	&	20260403	&	330	&	5.76	&	7.20	&	...	\\
$\alpha$ Tau	&	20260404	&	180	&	5.24	&	7.16	&	...	\\
$\beta$ Ori	&	20260404	&	180	&	5.13	&	7.15	&	...	\\
$\epsilon$ Sco	&	20260429	& 180	&	6.39	&	7.23	&	...	\\
$\delta$ Oph	&	20260430	&	180	&	7.12	&	7.28	&	...	\\
$\alpha$ Hya	&	20260501	&	240	&	5.23	&	7.39	&	...	\\
$\lambda$ Vel	&	20260501	&	120	&	5.68	&	7.22	&	...	\\
$\delta$ CMa	&	20260502	&	120	&	5.70	&	7.66	&	...	\\
\hline
\end{tabular}
\end{center}
\end{table}

\subsection{High-Contrast and Interferometric Vetting of Primary Reference Star Candidates}
In addition to moderate contrast vetting, we have executed hundreds of hours of observations between the 2024B-2026A semesters with high-contrast and interferometric facilities including VLT/SPHERE, LBT/SHARK-NIR, LBTI/LMIRCam, Subaru/SCExAO/CHARIS, Keck/NIRC2, VLTI/GRAVITY, VLTI/PIONIER, MagAO-X at Magellan-Clay, and CHARA/MIRC-X/MYSTIC (PIs: M. C. Vega-Pallauta, T. Mizuki, J. Hom, M. Brinjikji, N. Schragal, M. Millar-Blanchaer). In this work, we will report on the results from MagAO-X observations, while the results from our other high-contrast and interferometric programs will be reported in future publications of the CorGI-REx series (Vega-Pallauta in prep., Schragal in prep., Brinjikji in prep., Lallement in prep.). The results of these observations inform the re-assessment of rankings from CorGI-REx I \cite{hom2026}, as described in Section \ref{sec:rankings}.

\subsubsection{Reference Star Vetting with MagAO-X}
We used the extreme AO high-contrast imager MagAO-X\cite{males2024} to observe four reference star candidates--$\kappa$ Ori, $\gamma$ Ori, $\alpha$ Hya, and $\beta$ Crv over the 2025B and 2026A semesters (PIs: Schragal, Wang) in z'-band (908nm). As an additional technical demonstration for MagAO-X, we utilized the high-order wavefront control algorithm implicit electric field conjugation (iEFC\cite{haffert2023,haffert2026}) to obtain exceptionally deep contrasts. A knife-edge coronagraph was used, only allowing for a 180$^{\circ}$ FOV to be utilized per frame. Our observations are summarized in Table \ref{tab:magaoxObs}.

\begin{figure}
    \centering
    \includegraphics[width=0.95\linewidth]{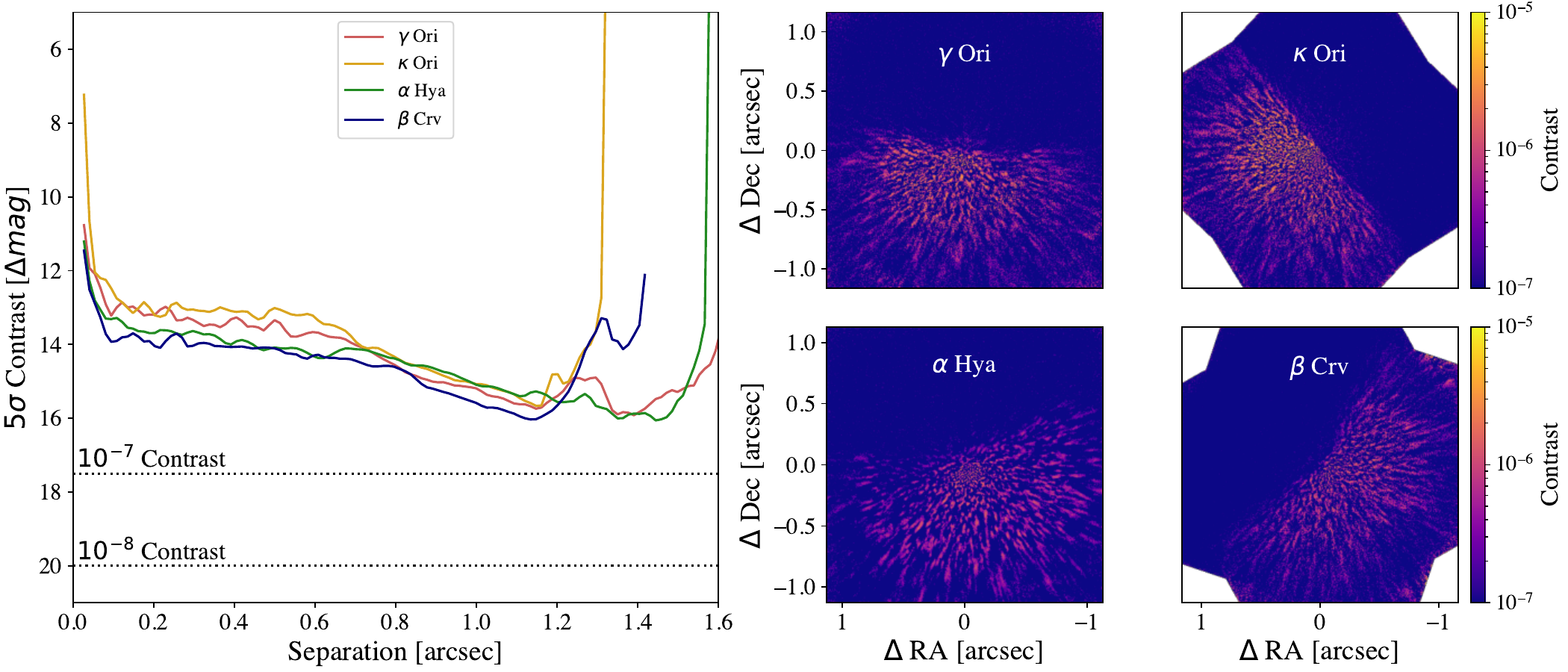}
    \caption{$5\sigma$ detection limits (left) and \texttt{pyklip}-reduced images (right) for the four MagAO-X targets observed. The reductions used 350 KL modes. These observations reach the deepest contrast limits possible (albeit only over a partial FOV) at wavelengths similar to Roman Coronagraph observing modes.}
    \label{fig:magaox}
\end{figure}

The images were dark-subtracted and registered using cross-correlation. Unocculted images of the host star were also taken to estimate the contrast of satellite spots to host star flux. This contrast conversion was applied to all of the frames. Out of the several thousands of frames at exposure times less than half a millisecond, only the few thousand with the highest SNR satellite spots were retained for the final reduction. We then coadded these frames into chunks grouped together by differences in parallactic angle less than $0.^{\circ}05$, leaving approximately 1000 frames per target for the final PSF subtraction. We used \texttt{pyklip}\cite{wang2015} with angular differential imaging (ADI), performing multiple reductions for each star with a varying range of KL modes applied from 1 to 1000. $5\sigma$ contrast curves were measured over the observable FOV, calibrated by injecting fake point sources into the dataset and measuring the algorithmic throughput loss through \texttt{pyklip}. Reduced images and measured contrast curves are shown in Figure \ref{fig:magaox}.

\begin{table}[ht]
\caption{Summary of reference star observations taken with MagAO-X.} 
\label{tab:magaoxObs}
\begin{center}       
\begin{tabular}{|l|l|l|l|l|}
\hline
Name	&	Date [UT]	&	$t_{int}$ [min]	&	$\Delta \theta$ [deg]	&	Seeing [$"$]	\\ \hline

$\gamma$ Ori	&	20251129	&	112	&	38.91	&	0."64	\\
$\kappa$ Ori	&	20251129	&	87	&	15.07	&	0."58	\\
$\alpha$ Hya	&	20260401	&	108	&	55.57	&	1."17	\\
$\beta$ Crv	&	20260526	&	100	&	42.49	&	0."69	\\
\hline
\end{tabular}
\end{center}
\end{table}

\section{Re-Assessment of Rankings from CorGI-REx I} \label{sec:rankings}
\begin{figure}
    \centering
    \includegraphics[width=0.75\linewidth]{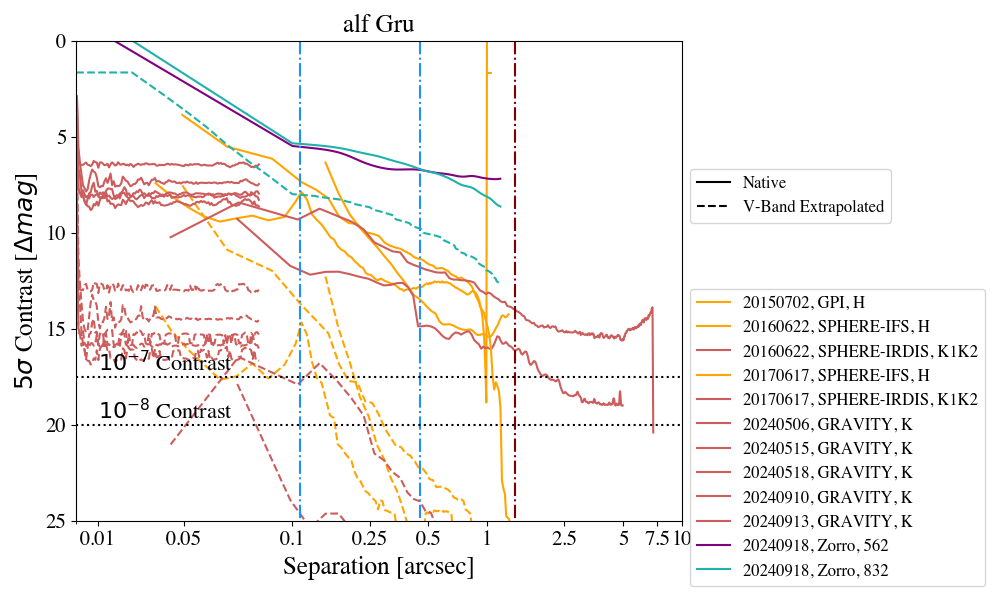}
    \caption{Collected detection limit curves for multiple observations of reference star candidate $\alpha$ Gru. The combination of detection limits from obtained and archival GPI, SPHERE, GRAVITY, and speckle interferometry observations maximally vet the star to limits approaching the performance levels of the Roman Coronagraph, encouraging its use in observations.}
    \label{fig:deep_vet}
\end{figure}

\begin{figure}
    \centering
    \includegraphics[width=0.75\linewidth]{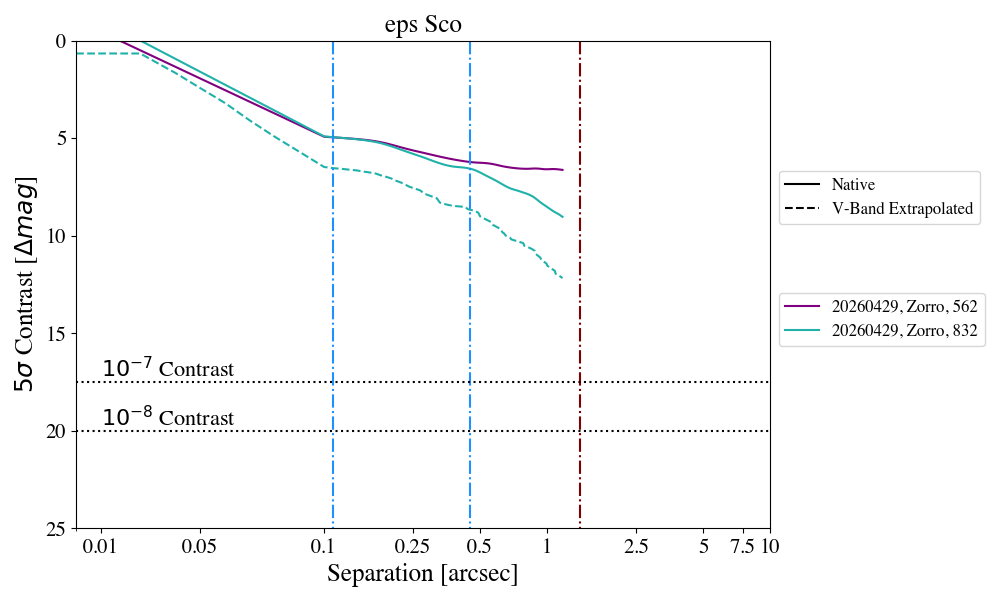}
    \caption{Collected detection limit curves for the reference star candidate $\epsilon$ Sco. Only speckle interferometry observations have been obtained, and therefore the star has only been vetted to a shallow contrast level. Fainter companions that could degrade Roman Coronagraph performance may still be present.}
    \label{fig:shallow_vet}
\end{figure}

Hom et al.\cite{hom2026} presented an initial list of 40 primary reference star candidates and 18 reserve reference star candidates and categorized confidence in suitability for Roman Coronagraph operations in the form of a ``rank", with confidence decreasing from ``A" to ``C." These rankings were assigned on the strength of evidence from literature, observational archives, and radial velocity/astrometry catalogs such as SB9\cite{pourbaix2004} and \textit{Gaia}. Hom et al.\cite{hom2026} presented the results of a shallow contrast speckle interferometry and adaptive optics imaging survey of these candidates, and the detection of any candidate companions would change these initial rankings. No candidates, however, were detected at high significance from the survey. Since the survey was completed, additional campaigns capable of reaching deeper contrast at different separation regimes (optical interferometry for separations $<100$mas and high-contrast imaging for separations $>100$mas) have identified several candidate companions and/or ruled out the presence of previously reported candidate companions. Depending on the properties of these companions, including whether or not they are confirmed, their corresponding ranks are adjusted. We collect as many relevant observations (from archival data and from our own observing programs) and measure corresponding detection limit curves as feasible to ultimately make a determination on changing the rank of each reference star. An example of a deeply-vetted reference star is shown in Figure \ref{fig:deep_vet} while an example of a shallow-vetted reference star is shown in Figure \ref{fig:shallow_vet}.

In addition to the presence of candidate companions, additional criteria have been introduced to reassess current reference star rankings. There are several factors that may upgrade or downgrade current reference star rankings. Reference star ranks may be upgraded if:
\begin{enumerate}
    \item Their data exhibit no candidate companions in one or more separation regime(s) of data
    \item Their data have deep detection limits from one or more separation regime(s) of data
    \item Their data achieve detection limit(s) that rule out a previously suggested candidate companion
\end{enumerate}

Reference star ranks may be downgraded if:
\begin{enumerate}
    \item They have an unconfirmed candidate companion
    \item They have a high near-IR excess indicative of hot dust\cite{ertel2025} but no indications of colder dust
    \item Their data are missing deep detection limits from one or more separation regime(s) of data
    \item There are lines of evidence that suggest a companion that have thus far not been ruled out by other means
    \item They have a low galactic latitude increasing the likelihood of a background object in the FOV
    \item For the HLC-NFOV observing mode only, have a stellar diameter larger than 2 mas.
    \item For the SPC-SPEC observing mode only, have a stellar diameter larger than 10 mas.
\end{enumerate}

Reference star candidates are entirely removed from consideration when:
\begin{enumerate}
    \item They have a confirmed companion that is expected to degrade Roman Coronagraph performance in the corresponding observing mode/contrast requirement
    \item They have substantial IR-excess ($\gtrsim10^{-5}$) from the near to far-IR
    \item They otherwise have some other confirmed property that is expected to degrade Roman Coronagraph performance in the corresponding observing mode/contrast requirement
\end{enumerate}

The presence of circumstellar dust is expected to introduce extended scattered light within the FOVs of the different Roman Coronagraph observing modes. This would negatively impact the efficacy of post-processing reference differential imaging (RDI), limiting achievable contrast. With new predictions on scattered-light contrasts from circumstellar dust, reference star candidates $\beta$ Leo and $\beta$ UMa were removed from consideration, with the predicted peak dust surface brightnesses approaching $10^{-7}$ contrast.

New candidate companions at separations $<2"$ have also been identified from recent high-contrast and interferometric imaging campaigns. While not all candidates have been confirmed, their presence ultimately affects their current rankings. The properties of these candidate companions will be reported in future publications.

Our re-ranking assessments for each instrument mode are provided in Table \ref{tab:re-rankings}. In addition to distinguishing rankings by mode, we also create distinct rankings for desired raw contrast scenarios (moderate vs. high) per mode, as some reference star properties may have a stronger impact on performance at high-contrast vs. moderate contrast.

\begin{table}[ht]
\caption{Version 2 rankings for reference star candidates, specified per mode (HLC-NFOV Band 1, SPC-SPEC Band 3, SPC-WFOV) and per desired contrast level (moderate and high).} 
\label{tab:re-rankings}
\begin{center}
\begin{tiny}
\begin{tabular}{|l|l|l|l|l|l|l|l|l|l|}
\hline
Star & $V$ & $I$ & v1 & v2 Rank & v2 Rank & v2 Rank & v2 Rank & v2 Rank & v2 Rank \\
& & & Rank & NFOV-High & NFOV-Mod & SPEC-High & SPEC-Mod & WFOV-High & WFOV-Mod \\
\hline
$\gamma$ Peg	&	2.83	&	3.31	&	C	&	A	&	A	&	A	&	A	&	A	&	A	\\
$\alpha$ Gru	&	1.73	&	1.95	&	B	&	A	&	A	&	A	&	A	&	A	&	A	\\
$\delta$ Leo	&	2.56	&	2.41	&	B	&	A	&	A	&	A	&	A	&	A	&	A	\\
$\eta$ UMa	&	1.85	&	2.12	&	B	&	A	&	A	&	A	&	A	&	A	&	A	\\
$\alpha$ Col	&	2.65	&	2.78	&	B	&	A/B	&	A/B	&	A/B	&	A/B	&	A/B	&	A/B	\\
$\beta$ Car	&	1.67	&	1.65	&	A	&	A/B	&	A/B	&	A/B	&	A/B	&	A/B	&	A/B	\\
$\epsilon$ Ori	&	1.7	&	1.93	&	B	&	A/B	&	A/B	&	A/B	&	A/B	&	A/B	&	A/B	\\
$\alpha$ Ara	&	2.84	&	3.29	&	B	&	B	&	B	&	B	&	B	&	B	&	B	\\
$\alpha$ Cep	&	2.45	&	2.24	&	B	&	B	&	B	&	B	&	B	&	B	&	B	\\
$\beta$ Lib	&	2.61	&	2.79	&	C	&	B	&	B	&	B	&	B	&	B	&	B	\\
$\beta$ TrA	&	2.83	&	2.43	&	B	&	B	&	B	&	B	&	B	&	B	&	B	\\
$\epsilon$ CMa	&	1.50	&	1.80	&	B	&	B	&	B	&	B	&	B	&	B	&	B	\\
$\eta$ Cen	&	2.33	&	2.67	&	B	&	B	&	B	&	B	&	B	&	B	&	B	\\
$\kappa$ Ori	&	2.06	&	2.27	&	A	&	B	&	B	&	B	&	B	&	B	&	B	\\
$\zeta$ Oph	&	2.56	&	2.50	&	C	&	B	&	B	&	B	&	B	&	B	&	B	\\
$\zeta$ Pup	&	2.25	&	2.58	&	C	&	B	&	B	&	F	&	B	&	F	&	C	\\
$\beta$ Tau	&	1.65	&	1.86	&	C	&	B	&	B	&	B	&	B	&	B	&	B	\\
$\gamma$ Ori	&	1.64	&	2.05	&	B	&	B	&	B	&	B	&	B	&	B	&	B	\\
$\rho$ Pup	&	2.83	&	2.37	&	B	&	B	&	B	&	B	&	B	&	B	&	B	\\
$\alpha$ Lep	&	2.58	&	2.14	&	B	&	B	&	B	&	B	&	B	&	B	&	B	\\
$\beta$ CMi*	&	2.89	&	3.07	&	B	&	B/C	&	B/C	&	B/C	&	B/C	&	B/C	&	B/C	\\
$\delta$ Cru	&	2.79	&	3.23	&	C	&	B/C	&	B/C	&	B/C	&	B/C	&	B/C	&	B/C	\\
$\delta$ Crv	&	2.94	&	3.03	&	C	&	B/C	&	B/C	&	B/C	&	B/C	&	B/C	&	B/C	\\
$\epsilon$ UMa	&	1.76	&	1.82	&	C	&	B/C	&	B/C	&	B/C	&	B/C	&	B/C	&	B/C	\\
$\alpha$ Peg	&	2.49	&	2.58	&	C	&	B/C	&	B/C	&	B/C	&	B/C	&	B/C	&	B/C	\\
$\beta$ CMa	&	1.98	&	2.35	&	A	&	B/C	&	B/C	&	B/C	&	B/C	&	B/C	&	B/C	\\
$\delta$ Cas	&	2.66	&	2.47	&	B	&	B/C	&	B/C	&	B/C	&	B/C	&	B/C	&	B/C	\\
$\gamma$ TrA	&	2.87	&	2.76	&	B	&	B/C	&	B/C	&	B/C	&	B/C	&	B/C	&	B/C	\\
$\alpha^2$ CVn	&	2.89	&	3.08	&	C	&	C	&	B	&	C	&	B	&	F	&	B	\\
$\zeta$ Aql	&	2.99	&	2.98	&	C	&	C	&	C	&	C	&	C	&	C	&	C	\\
$\beta$ Lup	&	2.68	&	3.08	&	B	&	C/F	&	C/F	&	C/F	&	C/F	&	C/F	&	C/F	\\
$\iota$ Car	&	2.21	&	1.88	&	C	&	F	&	F	&	F	&	F	&	F	&	F	\\
$\alpha$ Cyg	&	1.25	&	1.04	&	B	&	F	&	C	&	C	&	C	&	B	&	B	\\
$\alpha$ Hyi	&	2.86	&	2.41	&	C	&	F	&	F	&	F	&	F	&	F	&	F	\\
$\beta$ Cas	&	2.28	&	1.87	&	C	&	F	&	F	&	C	&	C	&	C	&	C	\\
$\beta$ Eri	&	2.78	&	2.57	&	C	&	F	&	F	&	F	&	F	&	F	&	F	\\
$\beta$ Leo	&	2.14	&	2.02	&	A	&	F	&	F	&	F	&	F	&	F	&	F	\\
$\beta$ UMa	&	2.34	&	2.36	&	B	&	F	&	F	&	F	&	F	&	F	&	F	\\
$\eta$ CMa	&	2.45	&	2.50	&	B	&	F	&	F	&	F	&	F	&	F	&	F	\\
$\eta$ Tau	&	2.85	&	2.88	&	C	&	F	&	B	&	B	&	B	&	B	&	B	\\
$\beta$ Hyi	&	2.79	&	1.94	&	B	&	F	&	B/C	&	C	&	B/C	&	B/C	&	B/C	\\
$\gamma$ Cyg	&	2.23	&	1.40	&	B	&	F	&	B	&	C	&	B	&	B	&	B	\\
$\beta$ Aqr	&	2.89	&	1.84	&	B	&	F	&	B/C	&	B/C	&	B/C	&	B/C	&	B/C	\\
$\beta$ Ori	&	0.13	&	0.15	&	A	&	F	&	B/C	&	C	&	B/C	&	B/C	&	B/C	\\
$\epsilon$ Leo	&	2.98	&	1.93	&	C	&	F	&	B/C	&	B/C	&	B/C	&	B/C	&	B/C	\\
$\beta$ Dra	&	2.81	&	1.64	&	C	&	F	&	B/C	&	B/C	&	B/C	&	B/C	&	B/C	\\
$\epsilon$ Gem	&	2.98	&	1.40	&	B	&	F	&	B	&	B	&	B	&	B	&	B	\\
$\alpha$ Aql	&	0.76	&	0.49	&	A	&	F	&	C	&	C	&	C	&	C	&	C	\\
$\alpha$ Per	&	1.79	&	...	&	C	&	F	&	B/C	&	B/C	&	B/C	&	B/C	&	B/C	\\
$\beta$ Crv	&	2.64	&	1.59	&	C	&	F	&	C	&	C	&	C	&	C	&	C	\\
$\alpha$ Aqr	&	2.94	&	1.76	&	C	&	F	&	C	&	C	&	C	&	C	&	C	\\
$\delta$ CMa	&	1.84	&	1.00	&	C	&	F	&	C	&	C	&	C	&	C	&	C	\\
$\epsilon$ Vir	&	2.79	&	1.71	&	B	&	F	&	B	&	B	&	B	&	B	&	B	\\
$\eta$ Dra	&	2.74	&	1.66	&	C	&	F	&	B	&	C	&	B	&	F	&	B	\\
$\gamma^2$ Sgr	&	2.99	&	1.75	&	C	&	F	&	B/C	&	B/C	&	B/C	&	B/C	&	B/C	\\
$\beta$ Oph	&	2.75	&	1.38	&	B	&	F	&	B	&	B	&	B	&	B	&	B	\\
$\lambda$ Sgr	&	2.81	&	1.50	&	B	&	F	&	B/C	&	B/C	&	B/C	&	B/C	&	B/C	\\
$\alpha$ Ser	&	2.63	&	1.25	&	B	&	F	&	B	&	B	&	B	&	B	&	B	\\
$\alpha$ Boo	&	-0.05	&	-1.68	&	...	&	F	&	F	&	F	&	F	&	B/C	&	B/C	\\
$\beta$ Gru	&	2.11	&	-1.57	&	...	&	F	&	F	&	C	&	B	&	A	&	A	\\
$\gamma$ Cru	&	1.64	&	-1.44	&	...	&	F	&	F	&	F	&	A	&	A	&	A	\\
$\alpha$ Tau	&	0.86	&	-1.31	&	...	&	F	&	F	&	F	&	F	&	B	&	B	\\
$\alpha$ Car	&	-0.74	&	-1.13	&	...	&	F	&	F	&	F	&	A/B	&	A/B	&	A/B	\\
$\beta$ Peg	&	2.42	&	-0.40	&	...	&	F	&	F	&	F	&	B/C	&	B/C	&	B/C	\\
$\beta$ And	&	2.05	&	-0.19	&	...	&	F	&	F	&	F	&	C	&	C	&	C	\\
$\beta$ Gem	&	1.14	&	-0.11	&	...	&	F	&	F	&	B	&	B	&	B	&	B	\\
$\mu$ Gem	&	2.87	&	-0.08	&	...	&	F	&	F	&	F	&	B/C	&	B/C	&	B/C	\\
$\rho$ Per	&	3.39	&	-0.03	&	...	&	F	&	F	&	F	&	B/C	&	B/C	&	B/C	\\
$\alpha$ Cet	&	2.53	&	0.02	&	...	&	F	&	F	&	F	&	A	&	A	&	A	\\
$\lambda$ Vel	&	2.21	&	0.03	&	...	&	F	&	F	&	F	&	A	&	A	&	A	\\
$\alpha$ Hya	&	1.97	&	0.16	&	...	&	F	&	F	&	A/B	&	A/B	&	A/B	&	A/B	\\
$\beta$ UMi	&	2.08	&	0.21	&	...	&	F	&	F	&	A	&	A	&	A	&	A	\\
$\gamma$ Dra	&	2.23	&	0.23	&	...	&	F	&	F	&	A	&	A	&	A	&	A	\\
$\delta$ Oph	&	2.75	&	0.43	&	...	&	F	&	F	&	F	&	F	&	F	&	F	\\
$\alpha$ Ari	&	2.01	&	0.54	&	...	&	F	&	F	&	B/C	&	B/C	&	B/C	&	B/C	\\
$\epsilon$ Peg	&	2.39	&	0.58	&	...	&	F	&	F	&	B	&	B	&	B	&	B	\\
$\theta$ Cen	&	2.05	&	0.76	&	...	&	F	&	F	&	A	&	A	&	A	&	A	\\
$\epsilon$ Sco	&	2.29	&	0.83	&	...	&	F	&	F	&	B	&	B	&	B	&	B	\\
$\alpha$ Cas	&	2.23	&	0.85	&	...	&	F	&	F	&	C	&	C	&	C	&	C	\\
\hline 
\end{tabular}
\end{tiny}
\end{center}
\end{table}

\section{Discussion} \label{sec:discussion}

\subsection{Scheduling Efficiency and Sky Coverage of Planned Science Targets}
\begin{figure}
    \centering
    \includegraphics[width=\linewidth]{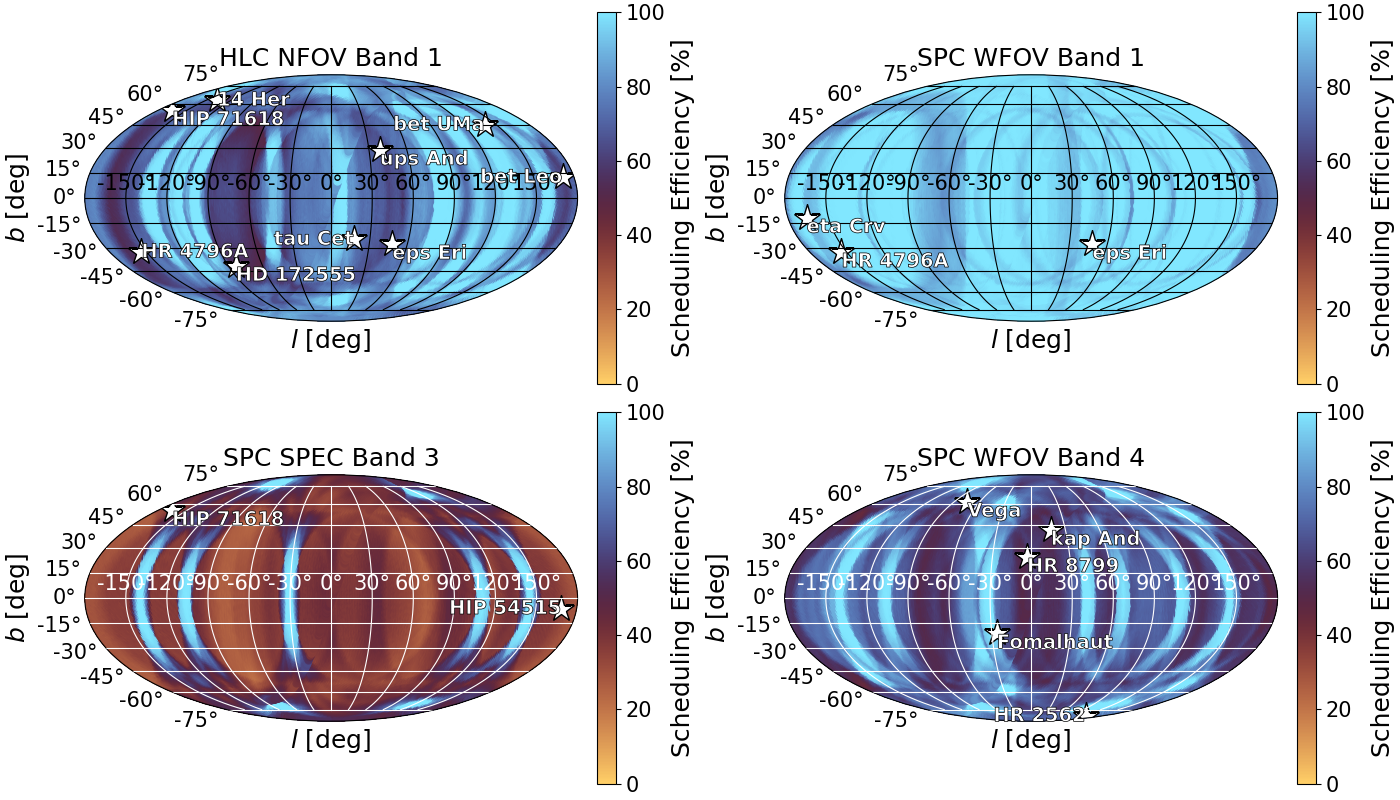}
    \caption{Scheduling efficiencies for high-contrast observations for the nominal four observing modes for the Roman Coronagraph in its first six months of observations. SPC-WFOV Band 1 observations are the easiest to schedule, as they lack any restriction on reference star diameter or target $I$-band magnitude. Scheduling efficiencies for HLC-NFOV Band 1 and SPC-WFOV Band 4 are somewhat similar. Overplotted are the sky positions of currently planned targets from the first six months of the Roman Coronagraph Observation Phase.}
    \label{fig:scheduling_efficiency}
\end{figure}

Hom et al.\cite{hom2026} showed that roughly 20 reference stars evenly distributed across the sky are needed for observing science targets $>50\%$ of the time they are visible from the observatory. The design and construction of observing campaigns occurs several months in advance of the actual observations, as sufficient time is needed to balance scheduling needs from both the Coronagraph Instrument and the Wide Field Imager. Given the limited amount of time allocated to the Coronagraph Instrument, only reference stars with the highest rankings (A-B) are selected for planning current observations. For the HLC-NFOV mode, there are currently 20 Rank A-B stars for high-contrast observations. This makes most programs able to be scheduled, but typically only in windows 3-7 days long per reference star option. As a result, while there are usually several possible windows under which specific campaigns can be scheduled, the windows are often short and do not continuously cover the full visibility window of the science target. Other observing modes (except for SPC-WFOV Band 1) have even more restrictive reference star options. In Table \ref{tab:efficiency}, we describe the scheduling efficiency\cite{hom2026}, which is the number of days in a year a given sky position has an available A- to B-ranked reference star within $\Delta$pitch$\leq5^{\circ}$ divided by the number of days in a year that sky position is visible from the observatory itself. We also show this visually for each observing mode in Figure \ref{fig:scheduling_efficiency} with currently planned science targets from the first six months of the Roman Coronagraph Observation Phase\cite{wolff2026}.

Aside from SPC-SPEC Band 3 observations, the current lists of Rank A-B reference stars for each mode allow for flexible and efficient scheduling of science programs, albeit with occasional short gaps in availability of reference stars within the $\Delta$pitch$\leq5^{\circ}$ criterion. These gaps can be overcome if the $\Delta$pitch restriction is slightly relaxed. While SPC-SPEC Band 3 observations are challenging to schedule, there are only two campaigns planned in the first six months of the Observation Phase. The lack of reference star options for this mode is primarily due to the mode having the most restrictive criteria for reference stars in addition to a lack of vetting observations. By the time planning for the next phase of Roman Coronagraph observations occur, more vetting observations are expected to be completed both with ground-based campaigns and vetting observations with the Roman Coronagraph itself on-sky\cite{wolff2026}.

\begin{table}[ht]
\caption{Number of reference stars, minimum scheduling efficiencies, and mean scheduling efficiencies for high-contrast observations for the nominal Roman Coronagraph observing modes assuming reference star ranks from A to B are suitable for observations. SPC-WFOV Band 1 observations are easiest to schedule due to the large number of possible reference stars. SPC-SPEC Band 3 observations are the most challenging to scheduling due to their strict reference star criteria. For moderate contrast observations per mode, the number of usable reference stars slightly increase and therefore slightly increase scheduling efficiency.} 
\label{tab:efficiency}
\begin{center}
\begin{tabular}{|l|l|l|l|}
\hline
Mode & Number of Ref. Stars & Minimum Efficiency & Mean Efficiency \\ \hline
HLC-NFOV Band 1 & 20 & 46.2\% & 80.4\% \\
SPC-WFOV Band 1 & 39 & 68.7\% & 96.2\% \\
SPC-SPEC Band 3 & 7 & 20.4\% & 47.2\% \\
SPC-WFOV Band 4 & 13 & 44.1\% & 72.0\% \\
\hline
\end{tabular}
\end{center}
\end{table}

\subsection{Impact of Reference Star Properties on Post-Processing Performance}
Due to the relatively lower amount of angular diversity from Roman Coronagraph observations (limited differential roll angle), post-processing of the different imaging modes of the Roman Coronagraph rely primarily on RDI. RDI is also preferred over ADI for post-processing images of circumstellar disks to avoid biases and artifacts from self-subtraction\cite{milli2012}. Post-processing, however, relies on the stability of the instrumental point spread function (PSF) between science images and a set of reference images. The poorer the correlation between speckle features of reference and science images, the worse the post-processing performance.

The presence of circumstellar dust and/or companions will contaminate the dark hole, possibly limiting achievable post-processing contrast. This will occur for companions both inside and outside of the FOV. If there are wide but bright companions, their PSF profile will leak light into the dark hole, while higher-contrast companions within the FOV will likely saturate reference star images, rendering them useless for RDI subtraction. 
Krist et al.\cite{krist2023} has also showed that the presence of jitter will impact the depth of the achievable contrast floor to varying degrees depending on the observing mode. Resolved stellar diameters can mimic this effect, motivating the stellar diameter criteria imposed for reference star selection. Resolved stellar diameters ultimately impact the performance of low-order wavefront sensing and control (LOWFSC) residuals, which ultimately impact the achievable raw contrast. Significant differences between science and reference target diameters therefore may impact achievable RDI post-processing performance, due to the nature of speckles being distinct between the two stars. 

To test the impacts of these properties, we utilize the Roman Coronagraph simulation package \texttt{CORGISIM}\cite{millar-blanchaer2024,zhang2026}. We simulate a set of fiducial observing scenarios, comparing the final post-processed images and achievable SNR of fiducial companions. For all simulated observing sequences, we assume two reference star visits and two science target visits at roll angles $+15^{\circ}$ and $-15^{\circ}$. Each simulation assumes HLC-NFOV Band 1 as the observing mode after achieving an average $5\times10^{-9}$ contrast across the FOV. The EXCAM detector settings for the reference star and science target were optimized to avoid saturation, achieve $SNR \geq 5$ on the faintest features in the FOV, and minimize the number of necessary frames to achieve that SNR. Integration time length was set for reaching $SNR = 5$ on a $5\times10^{-9}$ point source using \texttt{corgietc}. Target properties and EXCAM settings are described in Table \ref{tab:common_sim_setup}. For each fiducial observing scenario, we reduce the dataset with \texttt{corgidrp}\cite{millar-blanchaer2024}.  The simulations do not currently consider the time-varying behavior of the speckle field; therefore, absolute values of achieved contrast and point source SNR are overly optimistic. Regardless, the relative differences between achieved contrast and point source SNR between the different simulation cases convey the impact that different reference star properties have on achievable performance.

\begin{table}[ht]
\caption{Common inputs to \texttt{corgisim} across all simulated observing scenarios. The target properties were selected to be analogous with a reference and science star combination that may be used in the observation phase: $\beta$ Car (reference star) and $\epsilon$ Eri (science star).} 
\label{tab:common_sim_setup}
\begin{center}
\begin{tabular}{|l|l|l|}
\hline
\texttt{corgisim} input	&	Reference Star	&	Science Star	\\ \hline
Target $V$	&	1.69	&	3.73	\\
Spectral Type	&	A1V	&	K2V	\\
EM Gain	&	866.881	&	1728.718	\\
Frame Exposure Time [s]	&	22.55	&	49.931	\\
Total Integration Time [min]	&	23.3	&	69.9	\\
\hline
\end{tabular}
\end{center}
\end{table}

\subsubsection{Companions}
\begin{figure}
    \centering
    \includegraphics[width=\linewidth]{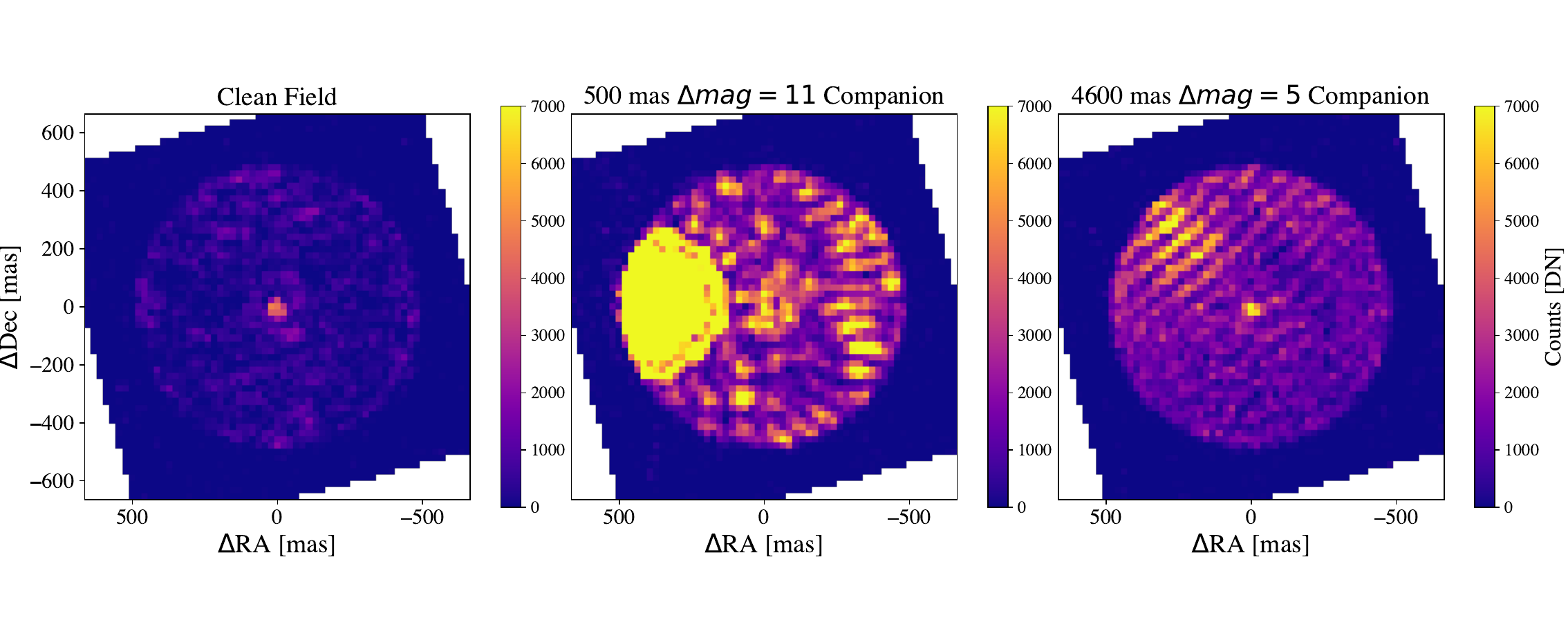}
    \caption{Comparison of reference star frames under three observing scenarios. (\textit{Left:}) No companions injected as part of the reference star scene. (\textit{Middle:}) $\Delta mag = 11$ companion injected at 500mas separation. Because the detector settings are optimized for high-contrast signals, the lower contrast companion severely saturates. (\textit{Right:}) $\Delta mag = 5.4$ companion injected at 4."6 separation. The PSF profile of the wide companion introduces light into the FOV but does not saturate.}
    \label{fig:sim_ref_frames}
\end{figure}

The presence of companions is one of the most important considerations when deciding whether or not a reference star is suitable. To demonstrate this, we generate a set of simulated observing sequences where companions of different brightnesses and separations are input into the reference star scenes. For the control simulation (``Clean Field"), we generate a simulated observation with no companions present in the reference star frames. For the first test case, we inject a $\Delta mag = 11$ companion at 500mas separation. This is representative of a candidate companion found around a formerly considered reference star. With the limited dynamic range of the detector and EM gain and exposure times optimized for $\sim10^{-9}$ contrasts, the companion and several of its airy rings saturate, rendering the reference frames unusable for RDI post-processing. For the second test case, we inject a $\Delta mag = 5.4$ companion at 4."6 separation. This scenario is an analogue of reference star candidate $\eta$ Dra, which is known to have a companion with these properties. In Figure \ref{fig:sim_ref_frames}, we show a derotated frame from each test. In the second test case, the PSF profile of the wide companion introduces light into the FOV but does not saturate.

\begin{figure}
    \centering
    \includegraphics[width=\linewidth]{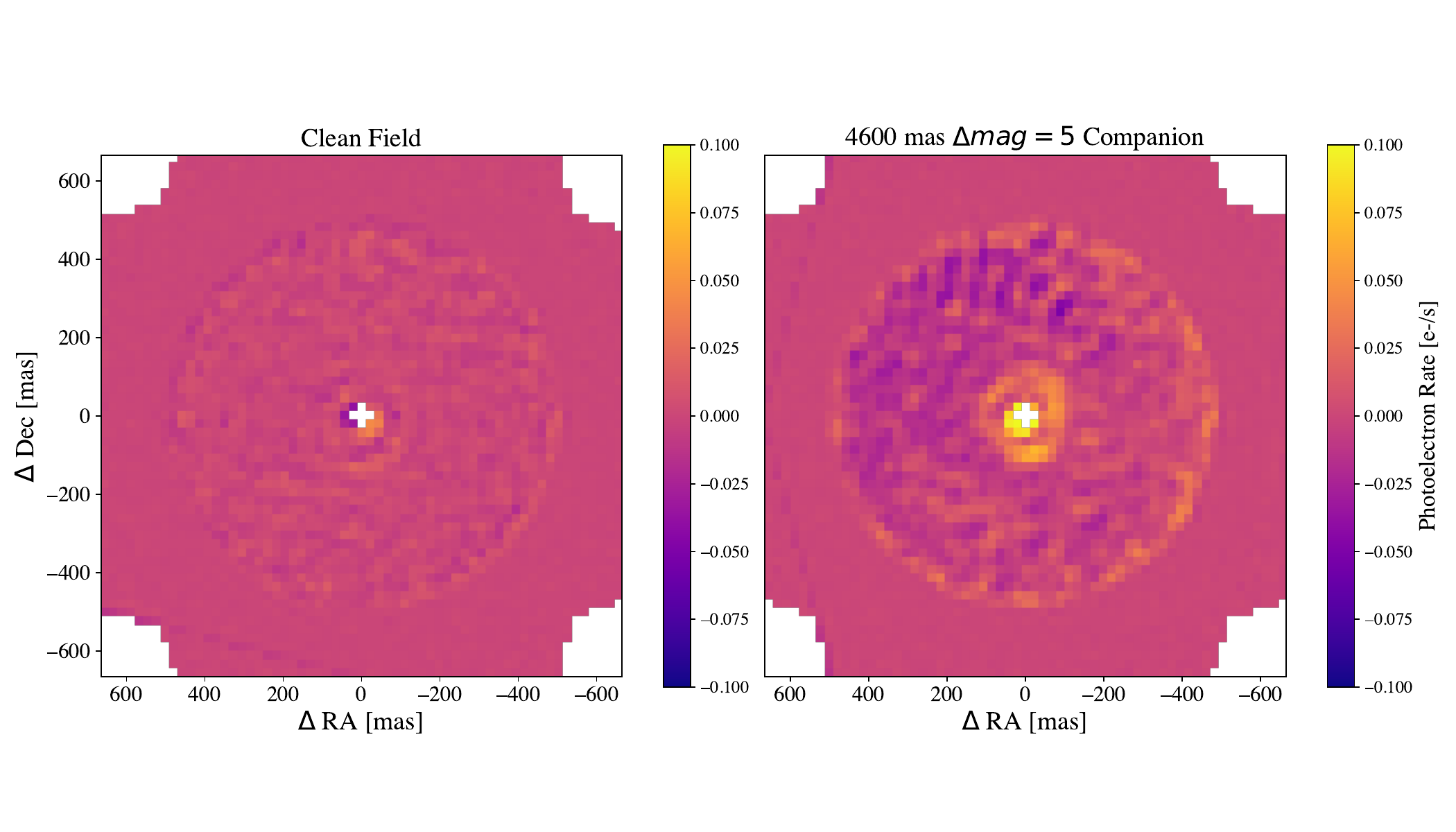}
    \caption{RDI reductions of the control observation scenario (``Clean Field", \textit{left}) and the wide companion scenario (\textit{right}). In the wide companion scenario, a region of the science image FOV is oversubtracted which would bias the $5\sigma$ contrast measurement.}
    \label{fig:sim_contrast_compare}
\end{figure}

\begin{figure}
    \centering
    \includegraphics[width=\linewidth]{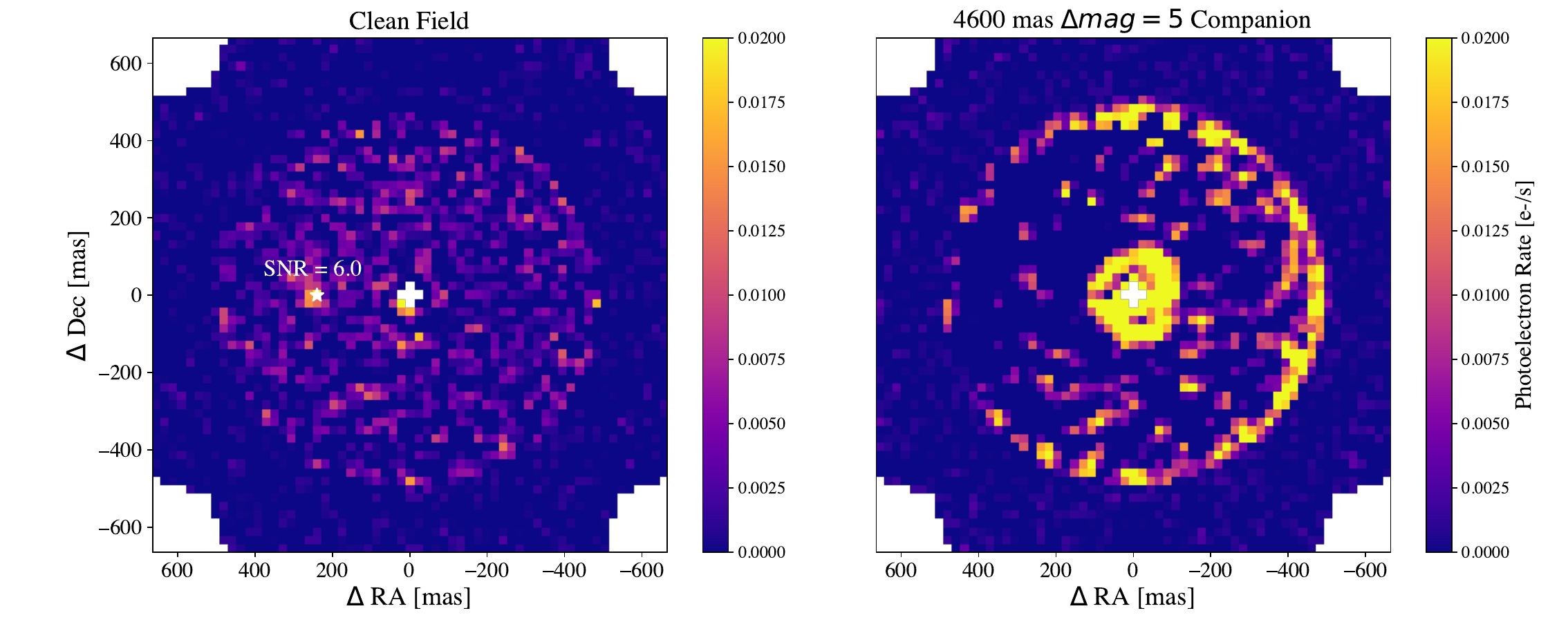}
    \caption{RDI reductions of the control observation scenario (``Clean Field", \textit{left}) and the wide companion scenario (\textit{right}) with an injected science target companion at 250mas separation. In the wide companion scenario, the oversubtraction of the science image attenuates the companion flux leading to SNR $<5$.}
    \label{fig:sim_snr_compare}
\end{figure}

To test impacts on RDI post-processing, we reduce the ``Clean Field" and 4."6 companion simulations, shown in Figure \ref{fig:sim_contrast_compare}. The PSF profile of the wide-companion biases the PSF model created from post-processing, resulting in an oversubtraction of signal in the final science image. This oversubtraction would also bias the measurements of any point-sources occupying the oversubtracted region along with the measured contrast. In Figure \ref{fig:sim_snr_compare}, we re-generated the ``Clean Field" and wide-companion observing scenarios with a $5\times10^{-9}$ companion at 250mas separation directly East of the science star. In the ``Clean Field" scenario, the companion is retrieved at $SNR>5$. In the wide-companion scenario, the oversubtraction of the wide-companion PSF profile severely attenuates the companion flux SNR to $SNR<5$. Lower-contrast companions are likely still detectable in the wide-companion scenario but their measured photometry may be biased.

\subsubsection{Resolved Stellar Diameters}
\begin{figure}
    \centering
    \includegraphics[width=\linewidth]{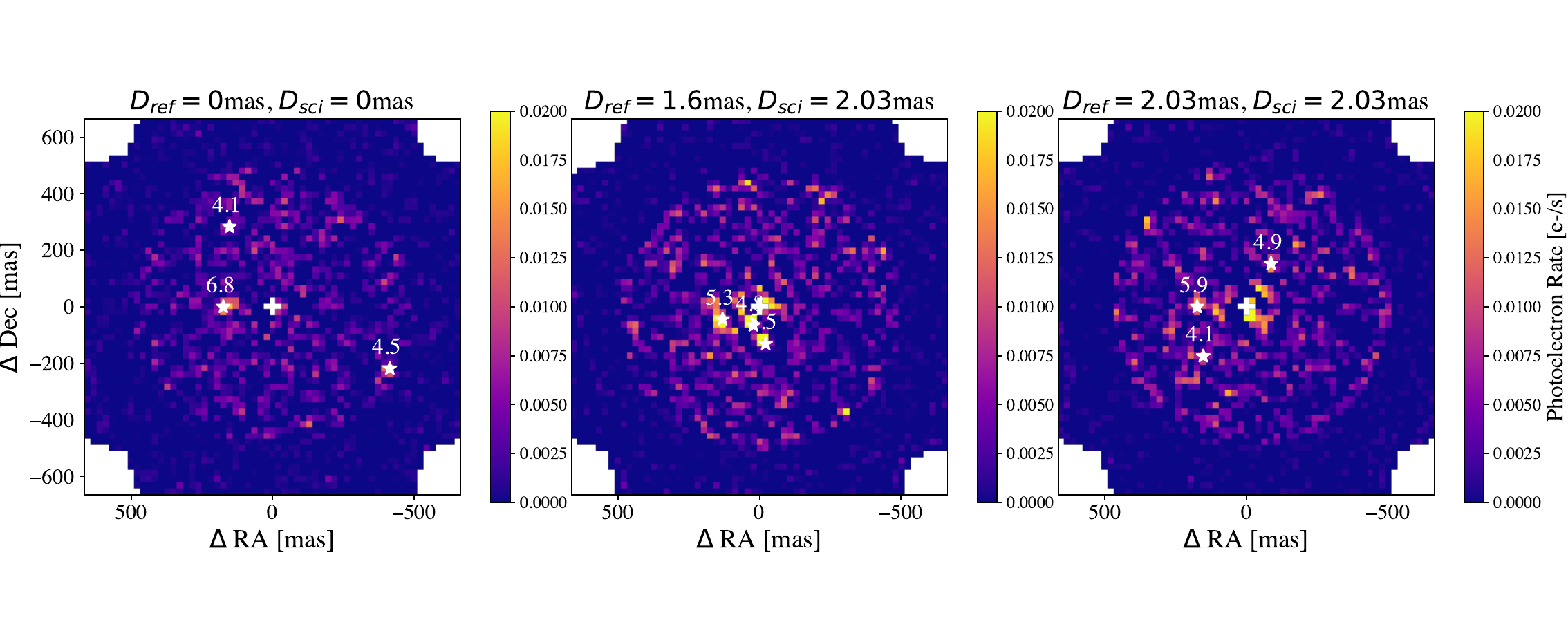}
    \caption{Comparison of a reduced simulated observing sequence assuming unresolved reference and science stars (\textit{left}, different resolved diameters for reference and science stars (\textit{middle}), and the same resolved diameters for reference and science stars (\textit{right}). The stars indicate the locations of point sources found in each image that achieve $SNR>4$. The simulated $5\times10^{-9}$ companion at 3.3$\lambda/D$ is retrieved from the left and right simulations and not in the middle simulated observing sequence. In the middle observing sequence, residuals originating from the larger resolved diameter of the science star persist and cannot be subtracted away with the reference star images.}
    \label{fig:diameter_compare}
\end{figure}

To test the impact of resolved stellar diameters on RDI post-processing performance in the HLC-NFOV mode, we simulate the effects of jitter with \texttt{corgisim}. Using the same setup as described in Table \ref{tab:common_sim_setup}, we place a companion at 3.3$\lambda / D$ ($\sim166$mas) East of the star, where contrast degradation from jitter/resolved stellar diameters are expected to be highest. We simulate three scenarios: 1. no implementation of resolved stellar diameter, 2. imposing diameters of 1.6mas and 2.03mas for the reference and science stars respectively, and 3. imposing diameters of 2.03mas for both the reference and science star. Our RDI-postprocessed images are shown in Figure \ref{fig:diameter_compare}. The companion is detectable in both the first and third cases, suggesting that even with the increased contrast degradation of observing a larger diameter science star, observing a similar diameter reference star could still allow for successful RDI postprocessing. The second scenario suggests differences of $\gtrsim0.4$mas diameter between the science and reference stars make detection of the companion challenging/impossible. It is important to note, however, that a limited set of post-processing reduction parameters were applied and not exhaustively explored, and that high-contrast companion detection may necessitate a need for other techniques such as companion forward-modeling.

\subsection{Implications for the Habitable Worlds Observatory}
The Roman Coronagraph is a critical technology pathfinder for a future Habitable Worlds Observatory coronagraph instrument. It will likely follow similar/related strategies to achieve even deeper contrasts, including the use of extremely bright reference stars and potentially RDI. If similar or even more strict criteria are adopted in addition to similar observational/analytic approaches, the instrument would likely use at least a subset of the reference stars described in this work. Various reference star properties, depending on their impact on overall performance, must therefore be considered in HWO coronagraph instrument design and operational structure. To minimize impact on HWO observational efficiency and performance, reference star criteria may need to be relaxed. If criteria cannot be relaxed, other observational and/or data reduction processes may need to be considered.

Relaxing intrinsic reference star criteria may expand the sample of usable HWO reference stars. For example, capability to observe fainter stars would significantly increase the reference star sample, but may not decrease overheads due to HOWFSC. While HWO will have a larger diameter and likely higher throughput, the deformable mirrors used for HOWFSC may have more modes to correct for, necessitating more reference star flux. Increasing or removing the stellar diameter limit is likely the most achievable criterion, given that performance degradation due to resolved stellar diameters depends on the coronagraph choice. Ultimately, expanding the diameter criterion does not significantly increase the sample of usable reference stars like increasing the magnitude limit would.

Improvements to instrument/observatory performance is likely another pathway that could be utilized to minimize impact to HWO observational efficiency. Hom et al.\cite{hom2026} showed that increasing the $\Delta$pitch restriction by a factor of two also expands the scheduling window for that specific reference/science pair by two. To minimize the impact of larger $\Delta$pitch angles, the observatory/instrument contrast would need to be robust against thermal variations. Additionally, efforts to increase dark hole stability over time would relax the frequency at which reference star observations may be needed.

Alternative approaches to wavefront sensing and control and/or data reduction techniques may also help to mitigate impacts on performance. For example, if there are companions present around a reference star candidate, angular differential imaging (ADI) could be used instead to post-process science data if there is sufficient angular diversity. For reference stars with exceptionally bright candidate companions, multi-star wavefront control\cite{thomas2015,sirbu2017} could be used.

\section{Summary} \label{sec:summary}
In this work, we have described the criteria used to construct a new reference star list specifically optimized for the SPC-SPEC and SPC-WFOV Roman Coronagraph observing modes. These stars must be extremely bright ($I<1.25$) and single. The current effort to vet reference star candidates for all Roman Coronagraph observing modes is on track to complete vetting of HLC-NFOV candidate reference stars down to contrasts of $\sim10^{-6}$ with powerful interferometric and high-contrast instrumentation. Simultaneously, efforts to begin vetting of SPC reference stars are underway using direct adaptive optics imaging, speckle interferometry, and extreme high-contrast wavefront control techniques with MagAO-X.

In tandem with obtained detection limits and newly detected candidate companions from all of these observing campaigns, we introduced additional criteria for ranking the suitability of reference star candidates to make updated decisions on their ultimate suitability for Roman Coronagraph observations, with quality ranks unique for each candidate, each observing mode, and two levels of desired contrast performance.

From these updated rankings, we determined the scheduling efficiencies over the course of a year for each observing mode. SPC-WFOV Band 1 observations are the easiest to schedule, as they have the loosest set of criteria. SPC-SPEC observations are the most challenging to schedule, given their extremely limiting criteria and lack of usable high-confidence reference stars.

To demonstrate the role of reference star characteristics on observation performance, we generated a series of simulated observing sequences to understand their impact on RDI post-processing performance. If lower-contrast companions are present in the FOV of reference star images, these images will likely become saturated and unusable for RDI post-processing. If there are wide-separation but bright companions, their PSF profile will introduce light into the FOV, biasing post-processing results and leading to oversubtraction in science images. Finally, if resolved stellar diameters are not well-matched between HLC-NFOV reference and science observations, achievable post-processing performance particularly near the IWA may be extremely limited.

Future work will involve the completion of ground-based surveys of reference star candidates and reporting the results of these surveys in future publications (Vega-Pallauta et al. in prep., Schragal et al. in prep., Brinjikji et al. in prep., Lallement et al. in prep.). Additional development of the simulations will also continue in an effort to add more fidelity to the results (e.g., introducing time-varying speckle behavior) and to investigate data reduction/processing approaches that can mitigate the impacts of reference star properties.

\appendix    

\acknowledgments 
The authors would also like to acknowledge the long list of astronomy community members, instrument principal investigators, and instrument team members who have volunteered time and resources into unofficial and official observations of CorGI-REx sample targets and observing proposal preparation. This list includes: Taichi Uyama, John D. Monnier, Rachael Roettenbacher, Jeremy Jones, Denis Mourard, Becky Flores, Steve Ertel, Jacopo Farinato, Valentina D'Orazi, Dino Mesa, Fernando Pedichini, Simone Antoniucci, Lucinda Lilley, Rebecca Zhang, Barnaby Norris, Julien Lozi, Thayne Currie, Connor Vancil, Jim Lyke, Eric Nielsen, Sloane Wiktorowicz, Jennifer Patience, Jarron Leisenring, Logan Moore, Jun Hashimoto, Jennifer Power, Jacob Isbell, Jared Carlson, John Debes, Katie Crotts, and Maria Vincent.

Based on observations obtained at the international Gemini Observatory, a program of NSF NOIRLab, which is managed by the Association of Universities for Research in Astronomy (AURA) under a cooperative agreement with the U.S. National Science Foundation on behalf of the Gemini Observatory partnership: the U.S. National Science Foundation (United States), National Research Council (Canada), Agencia Nacional de Investigaci\'{o}n y Desarrollo (Chile), Ministerio de Ciencia, Tecnolog\'{i}a e Innovaci\'{o}n (Argentina), Minist\'{e}rio da Ci\^{e}ncia, Tecnologia, Inova\c{c}\~{o}es e Comunica\c{c}\~{o}es (Brazil), and Korea Astronomy and Space Science Institute (Republic of Korea). Some of the observations in the paper made use of the High-Resolution Imaging instruments `Alopeke and Zorro. `Alopeke and Zorro were funded by the NASA Exoplanet Exploration Program and built at the NASA Ames Research Center by Steve B. Howell, Nic Scott, Elliott P. Horch, and Emmett Quigley. `Alopeke and Zorro were mounted on the Gemini North and South telescopes of the international Gemini Observatory, a program of NSF NOIRLab, which is managed by the Association of Universities for Research in Astronomy (AURA) under a cooperative agreement with the U.S. National Science Foundation. on behalf of the Gemini partnership: the U.S. National Science Foundation (United States), National Research Council (Canada), Agencia Nacional de Investigaci\'{o}n y Desarrollo (Chile), Ministerio de Ciencia, Tecnolog\'{i}a e Innovaci\'{o}n (Argentina), Minist\'{e}rio da Ci\^{e}ncia, Tecnologia, Inova\c{c}\~{o}es e Comunica\c{c}\~{o}es (Brazil), and Korea Astronomy and Space Science Institute (Republic of Korea). This work has made use of data from the European Space Agency (ESA) mission {\it Gaia} (\url{https://www.cosmos.esa.int/gaia}), processed by the {\it Gaia} Data Processing and Analysis Consortium (DPAC, \url{https://www.cosmos.esa.int/web/gaia/dpac/consortium}). Funding for the DPAC has been provided by national institutions, in particular the institutions participating in the {\it Gaia} Multilateral Agreement. Magellan @ Las Campanas Observatory access was supported by Northwestern University and the Center for Interdisciplinary Exploration and Research in Astrophysics (CIERA).

J.~R.~H. is funded by NASA under award No. 80NSSC25K0364. 
S.~B.~H. acknowledges support from the NASA Exoplanets Program Office. 
D.~R.~C. acknowledges partial support from NASA Grant 18-2XRP18\_2-0007. This research has made use of the Exoplanet Follow-up Observation Program (ExoFOP; DOI: 10.26134/ExoFOP5) website, which is operated by the California Institute of Technology, under contract with the National Aeronautics and Space Administration under the Exoplanet Exploration Program. Based on observations obtained at the Hale Telescope, Palomar Observatory, as part of a collaborative agreement between the Caltech Optical Observatories and the Jet Propulsion Laboratory operated by Caltech for NASA. 
D.~S. is funded by NASA under award No. 80NSSC24K0216. 
J.~J.~W. acknowledges support from NASA under award 80NSSC24K0087. 
Portions of this research were supported by funding from the Technology Research Initiative Fund (TRIF) of the Arizona Board of Regents and by generous anonymous philanthropic donations to the Steward Observatory of the College of Science at the University of Arizona. This research was carried out in part at the Jet Propulsion Laboratory, California Institute of Technology, under a contract with the National Aeronautics and Space Administration (80NM0018D0004). 
This material is based upon work supported by NASA under award Nos. 80NSSC24K0087, 80NSSC24K0097, 80NSSC24K0217, and 80NSSC25K0373. This work was co-authored by employees of Caltech/IPAC under Contract No. 80GSFC21R0032 with the National Aeronautics and Space Administration. 
We are very grateful for support from the NSF MRI Award No. 1625441. The Phase II upgrade program is made possible by the generous support of the Heising-Simons Foundation. MagAO-X uses the CACAO software package, which is supported by NSF Award No. 2410616.
 

\bibliography{report} 

@INPROCEEDINGS{bourges2014,
       author = {{Bourg{\'e}s}, L. and {Lafrasse}, S. and {Mella}, G. and {Chesneau}, O. and {Bouquin}, J.~L. and {Duvert}, G. and {Chelli}, A. and {Delfosse}, X.},
        title = "{The JMMC Stellar Diameters Catalog v2 (JSDC): A New Release Based on SearchCal Improvements}",
    booktitle = {Astronomical Data Analysis Software and Systems XXIII},
         year = 2014,
       editor = {{Manset}, N. and {Forshay}, P.},
       series = {Astronomical Society of the Pacific Conference Series},
       volume = {485},
        month = may,
        pages = {223},
       adsurl = {https://ui.adsabs.harvard.edu/abs/2014ASPC..485..223B}
}

@ARTICLE{pourbaix2004,
       author = {{Pourbaix}, D. and {Tokovinin}, A.~A. and {Batten}, A.~H. and {Fekel}, F.~C. and {Hartkopf}, W.~I. and {Levato}, H. and {Morrell}, N.~I. and {Torres}, G. and {Udry}, S.},
        title = "{SB9: The ninth catalogue of spectroscopic binary orbits}",
      journal = {Astronomy \& Astrophysics},
         year = 2004,
        month = sep,
       volume = {424},
        pages = {727-732},
          doi = {10.1051/0004-6361:20041213},
archivePrefix = {arXiv},
       eprint = {astro-ph/0406573},
 primaryClass = {astro-ph},
       adsurl = {https://ui.adsabs.harvard.edu/abs/2004A&A...424..727P}
}

@ARTICLE{krist2023,
       author = {{Krist}, John E. and {Steeves}, John B. and {Dube}, Brandon D. and {Eldorado Riggs}, A.~J. and {Kern}, Brian D. and {Marx}, David S. and {Cady}, Eric J. and {Zhou}, Hanying and {Poberezhskiy}, Ilya Y. and {Baker}, Caleb W. and {McGuire}, James P. and {Nemati}, Bijan and {Kuan}, Gary M. and {Mennesson}, Bertrand and {Trauger}, John T. and {Saini}, Navtej S. and {Rafels}, Sergi Hildebrandt},
        title = "{End-to-end numerical modeling of the Roman Space Telescope coronagraph}",
      journal = {Journal of Astronomical Telescopes, Instruments, and Systems},
         year = 2023,
        month = oct,
       volume = {9},
          eid = {045002},
        pages = {045002},
          doi = {10.1117/1.JATIS.9.4.045002},
archivePrefix = {arXiv},
       eprint = {2309.16012},
 primaryClass = {astro-ph.IM},
       adsurl = {https://ui.adsabs.harvard.edu/abs/2023JATIS...9d5002K}
}

@ARTICLE{mennesson2020,
       author = {{Mennesson}, B. and {Juanola-Parramon}, R. and {Nemati}, B. and {Ruane}, G. and {Bailey}, V.~P. and {Bolcar}, M. and {Martin}, S. and {Zimmerman}, N. and {Stark}, C. and {Pueyo}, L. and {Benford}, D. and {Cady}, E. and {Crill}, B. and {Douglas}, E. and {Gaudi}, B.~S. and {Kasdin}, J. and {Kern}, B. and {Krist}, J. and {Kruk}, J. and {Luchik}, T. and {Macintosh}, B. and {Mandell}, A. and {Mawet}, D. and {McEnery}, J. and {Meshkat}, T. and {Poberezhskiy}, I. and {Rhodes}, J. and {Riggs}, A.~J. and {Turnbull}, M. and {Roberge}, A. and {Shi}, F. and {Siegler}, N. and {Stapelfeldt}, K. and {Ygouf}, M. and {Zellem}, R. and {Zhao}, F.},
        title = "{Paving the Way to Future Missions: the Roman Space Telescope Coronagraph Technology Demonstration}",
      journal = {arXiv e-prints},
         year = 2020,
        month = aug,
          eid = {arXiv:2008.05624},
        pages = {arXiv:2008.05624},
          doi = {10.48550/arXiv.2008.05624},
archivePrefix = {arXiv},
       eprint = {2008.05624},
 primaryClass = {astro-ph.IM},
       adsurl = {https://ui.adsabs.harvard.edu/abs/2020arXiv200805624M}
}

@ARTICLE{scott2021,
       author = {{Scott}, Nicholas J. and {Howell}, Steve B. and {Gnilka}, Crystal L. and {Stephens}, Andrew W. and {Salinas}, Ricardo and {Matson}, Rachel A. and {Furlan}, Elise and {Horch}, Elliott P. and {Everett}, Mark E. and {Ciardi}, David R. and {Mills}, Dave and {Quigley}, Emmett A.},
        title = "{Twin High-resolution, High-speed Imagers for the Gemini Telescopes: Instrument description and science verification results}",
      journal = {Frontiers in Astronomy and Space Sciences},
         year = 2021,
        month = sep,
       volume = {8},
          eid = {138},
        pages = {138},
          doi = {10.3389/fspas.2021.716560},
       adsurl = {https://ui.adsabs.harvard.edu/abs/2021FrASS...8..138S}
}

@ARTICLE{hayward2001,
       author = {{Hayward}, T.~L. and {Brandl}, B. and {Pirger}, B. and {Blacken}, C. and {Gull}, G.~E. and {Schoenwald}, J. and {Houck}, J.~R.},
        title = "{PHARO: A Near-Infrared Camera for the Palomar Adaptive Optics System}",
      journal = {Publications of the Astronomy Society of the Pacific},
         year = 2001,
        month = jan,
       volume = {113},
       number = {779},
        pages = {105-118},
          doi = {10.1086/317969},
       adsurl = {https://ui.adsabs.harvard.edu/abs/2001PASP..113..105H}
}

@ARTICLE{milli2012,
       author = {{Milli}, J. and {Mouillet}, D. and {Lagrange}, A. -M. and {Boccaletti}, A. and {Mawet}, D. and {Chauvin}, G. and {Bonnefoy}, M.},
        title = "{Impact of angular differential imaging on circumstellar disk images}",
      journal = {Astronomy \& Astrophysics},
         year = 2012,
        month = sep,
       volume = {545},
          eid = {A111},
        pages = {A111},
          doi = {10.1051/0004-6361/201219687},
archivePrefix = {arXiv},
       eprint = {1207.5909},
 primaryClass = {astro-ph.EP},
       adsurl = {https://ui.adsabs.harvard.edu/abs/2012A&A...545A.111M}
}

@ARTICLE{thomas2015,
       author = {{Thomas}, S. and {Belikov}, R. and {Bendek}, E.},
        title = "{Techniques for High-contrast Imaging in Multi-star Systems. I. Super-Nyquist Wavefront Control}",
      journal = {The Astrophysical Journal},
         year = 2015,
        month = sep,
       volume = {810},
       number = {1},
          eid = {81},
        pages = {81},
          doi = {10.1088/0004-637X/810/1/81},
archivePrefix = {arXiv},
       eprint = {1501.01583},
 primaryClass = {astro-ph.IM},
       adsurl = {https://ui.adsabs.harvard.edu/abs/2015ApJ...810...81T}
}

@ARTICLE{sirbu2017,
       author = {{Sirbu}, D. and {Thomas}, S. and {Belikov}, R. and {Bendek}, E.},
        title = "{Techniques for High-contrast Imaging in Multi-star Systems. II. Multi-star Wavefront Control}",
      journal = {The Astrophysical Journal},
         year = 2017,
        month = nov,
       volume = {849},
       number = {2},
          eid = {142},
        pages = {142},
          doi = {10.3847/1538-4357/aa8e02},
archivePrefix = {arXiv},
       eprint = {1704.05441},
 primaryClass = {astro-ph.IM},
       adsurl = {https://ui.adsabs.harvard.edu/abs/2017ApJ...849..142S}
}

@ARTICLE{pecaut2013,
       author = {{Pecaut}, Mark J. and {Mamajek}, Eric E.},
        title = "{Intrinsic Colors, Temperatures, and Bolometric Corrections of Pre-main-sequence Stars}",
      journal = {The Astrophysical Journal Supplement Series},
         year = 2013,
        month = sep,
       volume = {208},
       number = {1},
          eid = {9},
        pages = {9},
          doi = {10.1088/0067-0049/208/1/9},
archivePrefix = {arXiv},
       eprint = {1307.2657},
 primaryClass = {astro-ph.SR},
       adsurl = {https://ui.adsabs.harvard.edu/abs/2013ApJS..208....9P}
}

@ARTICLE{horch2011,
       author = {{Horch}, Elliott P. and {Gomez}, Shamilia C. and {Sherry}, William H. and {Howell}, Steve B. and {Ciardi}, David R. and {Anderson}, Lisa M. and {van Altena}, William F.},
        title = "{Observations of Binary Stars with the Differential Speckle Survey Instrument. II. Hipparcos Stars Observed in 2010 January and June}",
      journal = {The Astronomical Journal},
         year = 2011,
        month = feb,
       volume = {141},
       number = {2},
          eid = {45},
        pages = {45},
          doi = {10.1088/0004-6256/141/2/45},
       adsurl = {https://ui.adsabs.harvard.edu/abs/2011AJ....141...45H}
}

@ARTICLE{howell2011,
       author = {{Howell}, Steve B. and {Everett}, Mark E. and {Sherry}, William and {Horch}, Elliott and {Ciardi}, David R.},
        title = "{Speckle Camera Observations for the NASA Kepler Mission Follow-up Program}",
      journal = {The Astronomical Journal},
         year = 2011,
        month = jul,
       volume = {142},
       number = {1},
          eid = {19},
        pages = {19},
          doi = {10.1088/0004-6256/142/1/19},
       adsurl = {https://ui.adsabs.harvard.edu/abs/2011AJ....142...19H}
}

@ARTICLE{gray2006,
       author = {{Gray}, R.~O. and {Corbally}, C.~J. and {Garrison}, R.~F. and {McFadden}, M.~T. and {Bubar}, E.~J. and {McGahee}, C.~E. and {O'Donoghue}, A.~A. and {Knox}, E.~R.},
        title = "{Contributions to the Nearby Stars (NStars) Project: Spectroscopy of Stars Earlier than M0 within 40 pc-The Southern Sample}",
      journal = {The Astronomical Journal},
         year = 2006,
        month = jul,
       volume = {132},
       number = {1},
        pages = {161-170},
          doi = {10.1086/504637},
archivePrefix = {arXiv},
       eprint = {astro-ph/0603770},
 primaryClass = {astro-ph},
       adsurl = {https://ui.adsabs.harvard.edu/abs/2006AJ....132..161G}
}

@ARTICLE{keenan1989,
       author = {{Keenan}, Philip C. and {McNeil}, Raymond C.},
        title = "{The Perkins Catalog of Revised MK Types for the Cooler Stars}",
      journal = {The Astrophysical Journal Supplement Series},
         year = 1989,
        month = oct,
       volume = {71},
        pages = {245},
          doi = {10.1086/191373},
       adsurl = {https://ui.adsabs.harvard.edu/abs/1989ApJS...71..245K}
}

@ARTICLE{gaiadr3,
       author = {{Gaia Collaboration} and {Vallenari}, A. and {Brown}, A.~G.~A. and {Prusti}, T. and {de Bruijne}, J.~H.~J. and {Arenou}, F. and {Babusiaux}, C. and {Biermann}, M. and {Creevey}, O.~L. and {Ducourant}, C. and {Evans}, D.~W. and {Eyer}, L. and {Guerra}, R. and {Hutton}, A. and {Jordi}, C. and {Klioner}, S.~A. and {Lammers}, U.~L. and {Lindegren}, L. and {Luri}, X. and {Mignard}, F. and {Panem}, C. and {Pourbaix}, D. and {Randich}, S. and {Sartoretti}, P. and {Soubiran}, C. and {Tanga}, P. and {Walton}, N.~A. and {Bailer-Jones}, C.~A.~L. and {Bastian}, U. and {Drimmel}, R. and {Jansen}, F. and {Katz}, D. and {Lattanzi}, M.~G. and {van Leeuwen}, F. and {Bakker}, J. and {Cacciari}, C. and {Casta{\~n}eda}, J. and {De Angeli}, F. and {Fabricius}, C. and {Fouesneau}, M. and {Fr{\'e}mat}, Y. and {Galluccio}, L. and {Guerrier}, A. and {Heiter}, U. and {Masana}, E. and {Messineo}, R. and {Mowlavi}, N. and {Nicolas}, C. and {Nienartowicz}, K. and {Pailler}, F. and {Panuzzo}, P. and {Riclet}, F. and {Roux}, W. and {Seabroke}, G.~M. and {Sordo}, R. and {Th{\'e}venin}, F. and {Gracia-Abril}, G. and {Portell}, J. and {Teyssier}, D. and {Altmann}, M. and {Andrae}, R. and {Audard}, M. and {Bellas-Velidis}, I. and {Benson}, K. and {Berthier}, J. and {Blomme}, R. and {Burgess}, P.~W. and {Busonero}, D. and {Busso}, G. and {C{\'a}novas}, H. and {Carry}, B. and {Cellino}, A. and {Cheek}, N. and {Clementini}, G. and {Damerdji}, Y. and {Davidson}, M. and {de Teodoro}, P. and {Nu{\~n}ez Campos}, M. and {Delchambre}, L. and {Dell'Oro}, A. and {Esquej}, P. and {Fern{\'a}ndez-Hern{\'a}ndez}, J. and {Fraile}, E. and {Garabato}, D. and {Garc{\'\i}a-Lario}, P. and {Gosset}, E. and {Haigron}, R. and {Halbwachs}, J. -L. and {Hambly}, N.~C. and {Harrison}, D.~L. and {Hern{\'a}ndez}, J. and {Hestroffer}, D. and {Hodgkin}, S.~T. and {Holl}, B. and {Jan{\ss}en}, K. and {Jevardat de Fombelle}, G. and {Jordan}, S. and {Krone-Martins}, A. and {Lanzafame}, A.~C. and {L{\"o}ffler}, W. and {Marchal}, O. and {Marrese}, P.~M. and {Moitinho}, A. and {Muinonen}, K. and {Osborne}, P. and {Pancino}, E. and {Pauwels}, T. and {Recio-Blanco}, A. and {Reyl{\'e}}, C. and {Riello}, M. and {Rimoldini}, L. and {Roegiers}, T. and {Rybizki}, J. and {Sarro}, L.~M. and {Siopis}, C. and {Smith}, M. and {Sozzetti}, A. and {Utrilla}, E. and {van Leeuwen}, M. and {Abbas}, U. and {{\'A}brah{\'a}m}, P. and {Abreu Aramburu}, A. and {Aerts}, C. and {Aguado}, J.~J. and {Ajaj}, M. and {Aldea-Montero}, F. and {Altavilla}, G. and {{\'A}lvarez}, M.~A. and {Alves}, J. and {Anders}, F. and {Anderson}, R.~I. and {Anglada Varela}, E. and {Antoja}, T. and {Baines}, D. and {Baker}, S.~G. and {Balaguer-N{\'u}{\~n}ez}, L. and {Balbinot}, E. and {Balog}, Z. and {Barache}, C. and {Barbato}, D. and {Barros}, M. and {Barstow}, M.~A. and {Bartolom{\'e}}, S. and {Bassilana}, J. -L. and {Bauchet}, N. and {Becciani}, U. and {Bellazzini}, M. and {Berihuete}, A. and {Bernet}, M. and {Bertone}, S. and {Bianchi}, L. and {Binnenfeld}, A. and {Blanco-Cuaresma}, S. and {Blazere}, A. and {Boch}, T. and {Bombrun}, A. and {Bossini}, D. and {Bouquillon}, S. and {Bragaglia}, A. and {Bramante}, L. and {Breedt}, E. and {Bressan}, A. and {Brouillet}, N. and {Brugaletta}, E. and {Bucciarelli}, B. and {Burlacu}, A. and {Butkevich}, A.~G. and {Buzzi}, R. and {Caffau}, E. and {Cancelliere}, R. and {Cantat-Gaudin}, T. and {Carballo}, R. and {Carlucci}, T. and {Carnerero}, M.~I. and {Carrasco}, J.~M. and {Casamiquela}, L. and {Castellani}, M. and {Castro-Ginard}, A. and {Chaoul}, L. and {Charlot}, P. and {Chemin}, L. and {Chiaramida}, V. and {Chiavassa}, A. and {Chornay}, N. and {Comoretto}, G. and {Contursi}, G. and {Cooper}, W.~J. and {Cornez}, T. and {Cowell}, S. and {Crifo}, F. and {Cropper}, M. and {Crosta}, M. and {Crowley}, C. and {Dafonte}, C. and {Dapergolas}, A. and {David}, M. and {David}, P. and {de Laverny}, P. and {De Luise}, F. and {De March}, R.},
        title = "{Gaia Data Release 3. Summary of the content and survey properties}",
      journal = {Astronomy \& Astrophysics},
         year = 2023,
        month = jun,
       volume = {674},
          eid = {A1},
        pages = {A1},
          doi = {10.1051/0004-6361/202243940},
archivePrefix = {arXiv},
       eprint = {2208.00211},
 primaryClass = {astro-ph.GA},
       adsurl = {https://ui.adsabs.harvard.edu/abs/2023A&A...674A...1G}
}

@ARTICLE{ertel2025,
       author = {{Ertel}, Steve and {Pearce}, Tim D. and {Debes}, John H. and {Faramaz}, Virginie C. and {Danchi}, William C. and {Anche}, Ramya M. and {Defr{\`e}re}, Denis and {Hasegawa}, Yasuhiro and {Hom}, Justin and {Kirchschlager}, Florian and {Rebollido}, Isabel and {Rousseau}, H{\'e}l{\`e}ne and {Scott}, Jeremy and {Stapelfeldt}, Karl and {Stuber}, Thomas A.},
        title = "{Review and Prospects of Hot Exozodiacal Dust Research For Future Exo-Earth Direct Imaging Missions}",
      journal = {Publications of the Astronomy Society of the Pacific},
         year = 2025,
        month = mar,
       volume = {137},
       number = {3},
          eid = {031001},
        pages = {031001},
          doi = {10.1088/1538-3873/adb6d5},
archivePrefix = {arXiv},
       eprint = {2504.00295},
 primaryClass = {astro-ph.EP},
       adsurl = {https://ui.adsabs.harvard.edu/abs/2025PASP..137c1001E}
}

@ARTICLE{hom2026,
       author = {{Hom}, Justin and {Wolff}, Schuyler G. and {Clark}, Catherine A. and {Ciardi}, David R. and {Deveny}, Sarah J. and {Howell}, Steve B. and {Greenbaum}, Alexandra Z. and {Littlefield}, Colin and {Anche}, Ramya M. and {Bailey}, Vanessa P. and {Brandner}, Wolfgang and {Chauvin}, Ga{\"e}l and {Girard}, Julien H. and {Kern}, Brian and {Mamajek}, Eric and {Mennesson}, Bertrand and {Savransky}, Dmitry and {Stapelfeldt}, Karl R. and {Biller}, Beth A. and {Brinjikji}, Marah and {Kuzuhara}, Masayuki and {Millar-Blanchaer}, Maxwell A. and {Mizuki}, Toshiyuki and {Schragal}, Nicholas T. and {Vega-Pallauta}, Macarena C. and {Wang}, Jason J. and {De Rosa}, Robert J. and {Douglas}, Ewan S. and {Macintosh}, Bruce and {Zhang}, Jingwen and {The Roman Coronagraph Community Participation Program}},
        title = "{CoronaGraph Instrument Reference Stars for Exoplanets (CorGI-REx). I. Preliminary Vetting and Implications for the Roman Coronagraph and Habitable Worlds Observatory}",
      journal = {The Astronomical Journal},
         year = 2026,
        month = jan,
       volume = {171},
       number = {1},
          eid = {36},
        pages = {36},
          doi = {10.3847/1538-3881/ae1d68},
archivePrefix = {arXiv},
       eprint = {2511.08862},
 primaryClass = {astro-ph.SR},
       adsurl = {https://ui.adsabs.harvard.edu/abs/2026AJ....171...36H}
}

@ARTICLE{badenes2018,
       author = {{Badenes}, Carles and {Mazzola}, Christine and {Thompson}, Todd A. and {Covey}, Kevin and {Freeman}, Peter E. and {Walker}, Matthew G. and {Moe}, Maxwell and {Troup}, Nicholas and {Nidever}, David and {Allende Prieto}, Carlos and {Andrews}, Brett and {Barb{\'a}}, Rodolfo H. and {Beers}, Timothy C. and {Bovy}, Jo and {Carlberg}, Joleen K. and {De Lee}, Nathan and {Johnson}, Jennifer and {Lewis}, Hannah and {Majewski}, Steven R. and {Pinsonneault}, Marc and {Sobeck}, Jennifer and {Stassun}, Keivan G. and {Stringfellow}, Guy S. and {Zasowski}, Gail},
        title = "{Stellar Multiplicity Meets Stellar Evolution and Metallicity: The APOGEE View}",
      journal = {The Astrophysical Journal},
         year = 2018,
        month = feb,
       volume = {854},
       number = {2},
          eid = {147},
        pages = {147},
          doi = {10.3847/1538-4357/aaa765},
archivePrefix = {arXiv},
       eprint = {1711.00660},
 primaryClass = {astro-ph.SR},
       adsurl = {https://ui.adsabs.harvard.edu/abs/2018ApJ...854..147B}
}

@INPROCEEDINGS{millar-blanchaer2024,
       author = {{Millar-Blanchaer}, Maxwell Andrew and {Wang}, Jason and {Bogat}, Ellis and {Schreiber}, J{\"u}rgen and {Ygouf}, Marie and {Ludwick}, Kevin J. and {Greenbaum}, Alexandra Z. and {Bailey}, Vanessa and {Hildebrandt Rafels}, Sergi and {Savransky}, Dmitry and {Samland}, Matthias and {Altinier}, Lisa and {Anche}, Ramya and {Biller}, Beth and {Chavez}, Amanda and {Choquet}, Elodie and {Girard}, Julien H. and {Hom}, Justin and {Ingalls}, James G. and {Kasdin}, N. Jeremy and {Krause}, Oliver and {Livingston}, John and {Mazoyer}, Johan and {Pueyo}, Laurent and {Uyama}, Taichi and {Zellem}, Robert T. and {Zimmerman}, Neil T.},
        title = "{The Roman coronagraph community participation program: data reduction and simulations}",
    booktitle = {Space Telescopes and Instrumentation 2024: Optical, Infrared, and Millimeter Wave},
         year = 2024,
       editor = {{Coyle}, Laura E. and {Matsuura}, Shuji and {Perrin}, Marshall D.},
       series = {Society of Photo-Optical Instrumentation Engineers (SPIE) Conference Series},
       volume = {13092},
        month = aug,
          eid = {1309256},
        pages = {1309256},
          doi = {10.1117/12.3020478},
       adsurl = {https://ui.adsabs.harvard.edu/abs/2024SPIE13092E..56M}
}

@ARTICLE{mozurkewich2003,
       author = {{Mozurkewich}, D. and {Armstrong}, J.~T. and {Hindsley}, R.~B. and {Quirrenbach}, A. and {Hummel}, C.~A. and {Hutter}, D.~J. and {Johnston}, K.~J. and {Hajian}, A.~R. and {Elias}, II, Nicholas M. and {Buscher}, D.~F. and {Simon}, R.~S.},
        title = "{Angular Diameters of Stars from the Mark III Optical Interferometer}",
      journal = {The Astronomical Journal},
         year = 2003,
        month = nov,
       volume = {126},
       number = {5},
        pages = {2502-2520},
          doi = {10.1086/378596},
       adsurl = {https://ui.adsabs.harvard.edu/abs/2003AJ....126.2502M}
}

@ARTICLE{gray1989F,
       author = {{Gray}, R.~O. and {Garrison}, R.~F.},
        title = "{The Early F-Type Stars: Refined Classification, Confrontation with Stroemgren Photometry, and the Effects of Rotation}",
      journal = {The Astrophysical Journal Supplement Series},
         year = 1989,
        month = feb,
       volume = {69},
        pages = {301},
          doi = {10.1086/191315},
       adsurl = {https://ui.adsabs.harvard.edu/abs/1989ApJS...69..301G}
}

@INPROCEEDINGS{males2024,
       author = {{Males}, Jared R. and {Close}, Laird M. and {Haffert}, Sebastiaan Y. and {Kautz}, Maggie Y. and {Kueny}, Jay and {Long}, Joseph D. and {McEwen}, Eden and {Swimmer}, Noah and {Bailey}, John I. and {Foster}, Warren and {Mazin}, Benjamin A. and {Pearce}, Logan and {Liberman}, Joshua and {Twitchell}, Katie and {Weinberger}, Alycia J. and {Guyon}, Olivier and {Hedglen}, Alexander D. and {McLeod}, Avalon and {Roberts}, Roz and {Van Gorkom}, Kyle and {Li}, Jialin and {Doty}, Isabella and {Gasho}, Victor},
        title = "{MagAO-X: commissioning results and status of ongoing upgrades}",
    booktitle = {Adaptive Optics Systems IX},
         year = 2024,
       editor = {{Jackson}, Kathryn J. and {Schmidt}, Dirk and {Vernet}, Elise},
       series = {Society of Photo-Optical Instrumentation Engineers (SPIE) Conference Series},
       volume = {13097},
        month = aug,
          eid = {1309709},
        pages = {1309709},
          doi = {10.1117/12.3019464},
       adsurl = {https://ui.adsabs.harvard.edu/abs/2024SPIE13097E..09M}
}

@software{wang2015,
       author = {{Wang}, Jason J. and {Ruffio}, Jean-Baptise and {De Rosa}, Robert J. and {Aguilar}, Jonathan and {Wolff}, Schuyler G. and {Pueyo}, Laurent},
        title = "{pyKLIP: PSF Subtraction for Exoplanets and Disks}",
 howpublished = {Astrophysics Source Code Library, record ascl:1506.001},
         year = 2015,
        month = jun,
          eid = {ascl:1506.001},
archivePrefix = {ascl},
       eprint = {1506.001},
       adsurl = {https://ui.adsabs.harvard.edu/abs/2015ascl.soft06001W}
}

@ARTICLE{haffert2023,
       author = {{Haffert}, S.~Y. and {Males}, J.~R. and {Ahn}, K. and {Van Gorkom}, K. and {Guyon}, O. and {Close}, L.~M. and {Long}, J.~D. and {Hedglen}, A.~D. and {Schatz}, L. and {Kautz}, M. and {Lumbres}, J. and {Rodack}, A. and {Knight}, J.~M. and {Miller}, K.},
        title = "{Implicit electric field conjugation: Data-driven focal plane control}",
      journal = {Astronomy \& Astrophysics},
         year = 2023,
        month = may,
       volume = {673},
          eid = {A28},
        pages = {A28},
          doi = {10.1051/0004-6361/202244960},
archivePrefix = {arXiv},
       eprint = {2303.13719},
 primaryClass = {astro-ph.IM},
       adsurl = {https://ui.adsabs.harvard.edu/abs/2023A&A...673A..28H}
}

@INPROCEEDINGS{wolff2026,
    author = {{Wolff}, Schuyler G. },
    title = "{The Roman Coronagraph Community Participation Program: trials and triumphs of designing an observing program for a technology demonstration instrument}",
    booktitle = {Space Telescopes and Instrumentation 2026: Optical, Infrared, and Millimeter Wave},
year = 2026,
series = {Society of Photo-Optical Instrumentation Engineers (SPIE) Conference Series}
}

@INPROCEEDINGS{zhang2026,
    author = {{Zhang}, Jingwen. },
    title = "{The Roman Coronagraph Community Participation Program: “corgisim” — a simulation suite for the Nancy Grace Roman Space Telescope Coronagraphic Instrument}",
    booktitle = {Space Telescopes and Instrumentation 2026: Optical, Infrared, and Millimeter Wave},
year = 2026,
series = {Society of Photo-Optical Instrumentation Engineers (SPIE) Conference Series}
}

@ARTICLE{haffert2026,
       author = {{Haffert}, S.~Y. and {Liberman}, J. and {Males}, J.~R. and {Close}, L.~M. and {Foster}, W.~B. and {Van Gorkom}, K. and {Guyon}, O. and {Hedglen}, A.~D. and {Johnson}, P.~T. and {Kautz}, M.~Y. and {Kueny}, J.~K. and {Li}, J. and {Long}, J.~D. and {Lumbres}, J. and {Mars}, M. and {McEwen}, E.~A. and {McLeod}, A. and {Schatz}, L. and {Tonucci}, E. and {Twitchell}, K.},
        title = "{On-sky dark hole diggin' with implicit Electric Field Conjugation on MagAO-X}",
      journal = {arXiv e-prints},
         year = 2026,
        month = jul,
          eid = {arXiv:2607.08146},
        pages = {arXiv:2607.08146},
          doi = {10.48550/arXiv.2607.08146},
archivePrefix = {arXiv},
       eprint = {2607.08146},
 primaryClass = {astro-ph.IM},
       adsurl = {https://ui.adsabs.harvard.edu/abs/2026arXiv260708146H}
}
\bibliographystyle{spiebib} 

\end{document}